\documentclass[aps]{revtex4}

\def\ov#1{\overline{#1}}

\def\wt#1{\widetilde{#1}}
\def\vb#1{\mbox{\boldmath$#1$}}
\def\pd#1#2{\frac{\partial #1}{\partial #2}}

\def\wh#1{\widehat{#1}}
\def\bdot{\,\vb{\cdot}\,}
\def\btimes{\,\vb{\times}\,}

\def\bhat{\wh{{\sf b}}}
\def\cal#1{\mathcal{#1}}

\def\bhat{\wh{{\sf b}}}

\newcommand{\bc}{\begin{center}}
\newcommand{\ec}{\end{center}}
\newcommand{\bt}{\begin{tabbing}}
\newcommand{\et}{\end{tabbing}} 
\newcommand{\be}{\begin{equation}}
\newcommand{\ee}{\end{equation}}
\newcommand{\ba}{\begin{eqnarray}}
\newcommand{\ea}{\end{eqnarray}}

\newcommand{\gav}[1]{
\langle #1 \rangle
}

\begin{document}

\title{Guiding-center recursive Vlasov and Lie-transform methods in plasma physics}

\author{A.\,J. Brizard$^1$ and A.\,Mishchenko$^2$}
\affiliation{$^1$Department of Physics, Saint Michael's College, Colchester, VT 05439, USA \\
$^2$Max-Planck-Institut f\"{u}r Plasmaphysik, EURATOM-Association, D-17491, Greifswald, Germany}

\begin{abstract}
The gyrocenter phase-space transformation used to describe nonlinear gyrokinetic theory is rediscovered by a recursive solution of the Hamiltonian dynamics associated with the perturbed guiding-center Vlasov operator. The present work clarifies the relation between the derivation of the gyrocenter phase-space coordinates by the guiding-center recursive Vlasov method and the method of Lie-transform phase-space transformations. 
\end{abstract}

\maketitle

\section{Introduction}

A common technique \cite{Davidson,HTH} associated with the multiple space-time-scale solution of the Vlasov equation 
\begin{equation}
0 \;=\; \frac{df}{dt} \;=\; \pd{f}{t} \;+\; \left( v_{\|}\,\bhat + {\bf v}_{\bot}\right)\bdot\nabla f \;+\; \frac{q}{m} \left( {\bf E} \;+\; 
\frac{{\bf v}}{c}\btimes{\bf B} \right)\bdot\pd{f}{{\bf v}}
\label{eq:Vlasov_eq}
\end{equation}
is to expand the Vlasov operator $L \equiv d/dt = L_{0} + \epsilon\,L_{1} + \epsilon^{2}\,L_{2} + \cdots$ and the Vlasov distribution $f \equiv f_{0} + \epsilon\,f_{1}+ \cdots$ asymptotically in powers of a small ordering parameter $\epsilon$. One then proceeds with a recursive solution of the Vlasov equation (\ref{eq:Vlasov_eq}) at each order in the hierarchy:
\begin{equation}
\left. \begin{array}{rcl}
0 & = & L_{0}f_{0} \\
0 & = & L_{1}f_{0} \;+\; L_{0}f_{1} \\
0 & = & L_{2}f_{0} \;+\; L_{1}f_{1} \;+\; L_{0}f_{2} \\
 & \vdots & 
\end{array} \right\}.
\label{eq:Vlasov_hierarchy}
\end{equation}
Two important assumptions are associated with the hierarchy (\ref{eq:Vlasov_hierarchy}). First, we assume that the lowest-order equation 
$L_{0}f_{0} = 0$ has a known (exact) solution. For example, the lowest-order dynamics characterized by $L_{0}$ is often associated with a cyclic 
(lowest-order) orbital angle $\varphi$ (i.e., $L_{0} \equiv \omega_{\varphi}\,\partial/\partial\varphi$, where $\omega_{\varphi} \equiv d\varphi/dt$) so that the lowest-order Vlasov equation $L_{0}f_{0} \equiv 0$ simply implies that $f_{0}$ is independent of the orbital angle $\varphi$. Second, we assume that the operator $L_{0}$ can be inverted (e.g., $L_{0}^{-1}f \equiv \omega_{\varphi}^{-1}\,\int f\,d\varphi$) so that the solution for the first-order correction $f_{1}$ in (\ref{eq:Vlasov_hierarchy}) may be written as $f_{1} = \ov{f}_{1} - L_{0}^{-1}(L_{1}f_{0})$, where $\ov{f}_{1}$ satisfies the homogeneous equation $L_{0}\ov{f}_{1} \equiv 0$. The small ordering parameter $\epsilon$ appearing in this asymptotic expansion is defined by the relation $L_{0}^{-1}\,L_{k} \equiv {\cal O}(\epsilon^{k})$.

The purpose of the present paper is to compare and contrast the derivation of gyrocenter phase-space coordinates by the guiding-center recursive Vlasov method and the method of Lie-transform phase-space transformations. In \S\,\ref{sec:Vlasov} the expansion of the Vlasov operator $L = d/dt$ defined in (\ref{eq:Vlasov_eq}) is given in powers of a small ordering parameter $\epsilon$. The guiding-center recursive Vlasov (gcrV) method is introduced in \S\,\ref{sec:gcrv} based on expansion of the guiding-center Vlasov operator $L_{\rm gc} \equiv {\sf T}_{\rm gc}^{-1}L{\sf T}_{\rm gc}$ defined in terms of the guiding-center push-forward $({\sf T}_{\rm gc}^{-1})$ and pull-back $({\sf T}_{\rm gc})$ operators. In order to keep the analysis focussed on gyrokinetic applications, we assume that the background magnetic field is uniform and that the particles move under the influence of a fluctuating electrostatic field (with slow and fast space-time scales). Using the gcrV method, we derive explicit expressions for the gyrocenter phase-space coordinates. In \S\,\ref{sec:gyro_Lie}, the gyrocenter Lie-Transform (gyLt) method is applied to the derivation of the gyrocenter phase-space coordinates. In order to allow comparison with expressions derived by the gcrV method, higher-order terms are kept. In \S\,\ref{sec:gyro_Vlasov_eq}, the gyrokinetic Vlasov equation is derived both by the gcrV and gyLt methods. While the gcrV derivation does not allow for a systematic truncation scheme that preserves energy conservation, the gyLt method does since it is naturally associated with a variational formulation. We also introduce the gyrocenter pull-back operator ${\sf T}_{\rm gy}$ and discuss its physical interpretation. In \S\,\ref{sec:Poisson}, we present the gyrokinetic Poisson equation expressed in terms of the gyrocenter moment (with respect to the gyrocenter Vlasov distribution) of the gyrocenter push-forward 
${\sf T}_{\rm gy}^{-1}\delta_{\rm gc}^{3}$ of the guiding-center delta function $\delta_{\rm gc}^{3} \equiv {\sf T}_{\rm gc}^{-1}\delta^{3}({\bf x} - 
{\bf r})$. Lastly, our work is summarized in \S\,\ref{sec:sum} and Appendix A presents the guiding-center phase-space transformation for nonuniform magnetic fields.

\section{\label{sec:Vlasov}Expansion of the Vlasov Operator}

We begin with the expansion of the Vlasov operator
\begin{equation}
\frac{d}{dt} \;=\; \pd{}{t} \;+\; \left( v_{\|}\,\bhat + {\bf v}_{\bot}\right)\bdot\nabla \;+\; \frac{q}{m} \left( {\bf E} \;+\; \frac{{\bf v}}{c}
\btimes{\bf B} \right)\bdot\pd{}{{\bf v}} \;\equiv\; L_{0} \;+\; \epsilon\;L_{1},
\label{eq:Vlasov_L}
\end{equation}
which is written in terms of the two operators $L_{0}$ and $L_{1}$ (defined below). Here, within the context of the derivation of gyrocenter coordinates for an electrostatic gyrokinetic model \cite{Dubin}, we consider the case of a strong background magnetic field and a small quasi-static electric field ${\bf E} = -\,\nabla\phi$. To simplify the analysis, we also consider a uniform background magnetic field (i.e, the unperturbed guiding-center dynamics is represented simply in terms of parallel motion along straight magnetic field lines).

\subsection{Zeroth-order Vlasov Operator}

In a uniform magnetic field, the zeroth-order Vlasov operator $L_{0}$ is defined as
\begin{equation}
L_{0}G \;\equiv\; \Omega\;\left( \pd{}{\theta} \;+\; \pd{\vb{\rho}}{\theta}\bdot\nabla \right) G \;\equiv\; \Omega\;{\sf T}_{{\rm gc}}\left[
\pd{}{\theta}\left( {\sf T}_{{\rm gc}}^{-1}G \frac{}{}\right)\right],
\label{eq:L0_def}
\end{equation}
where $G$ is an arbitrary function on particle phase space and the perpendicular velocity ${\bf v}_{\bot} \equiv \Omega\,\partial\vb{\rho}_{\rm gc}/\partial\theta$ is expressed in terms of the gyroradius vector $\vb{\rho}_{\rm gc} = (\bhat/\Omega)\btimes{\bf v}$, which depends explicitly on the gyroangle $\theta$. In the second expression in (\ref{eq:L0_def}), we introduced the guiding-center pull-back ${\sf T}_{{\rm gc}} \equiv  \exp(-\vb{\rho}_{\rm gc}\bdot\nabla)$ and the guiding-center push-forward ${\sf T}_{{\rm gc}}^{-1} \equiv \exp(\vb{\rho}_{\rm gc}\bdot\nabla)$, which are both associated with the guiding-center phase-space transformation for a uniform magnetic field (the guiding-center pull-back and push-forward operators are given for a nonuniform magnetic field in Appendix A). Note that the second term $({\bf v}_{\bot}\bdot\nabla)$ in (\ref{eq:L0_def}) may be of the same order of magnitude as the first term $(\Omega\,\partial/
\partial\theta)$ when it is applied to short-wavelength fluctuating fields (e.g., fluctuations that satisfy the gyrokinetic ordering 
\cite{Brizard_Hahm}).

After defining the guiding-center push-forward of an arbitrary function $G$ on particle phase space:
\begin{equation}
G_{{\rm gc}} \;\equiv\; {\sf T}_{{\rm gc}}^{-1}G,
\label{eq:Fgc_def}
\end{equation}
we arrive at the final expression for the zeroth-order Vlasov operator:
\begin{equation}
L_{0}G \;=\; {\sf T}_{{\rm gc}}\left( \Omega\;\pd{G_{{\rm gc}}}{\theta} \right) \;\equiv\; {\sf T}_{{\rm gc}}\left(\, L_{0{\rm gc}}\,G_{{\rm gc}}
\frac{}{}\right),
\label{eq:L0_gc}
\end{equation}
where $L_{0{\rm gc}} \equiv \Omega\,\partial/\partial\theta$ is the lowest-order guiding-center Vlasov operator. This operator can easily be inverted:
\begin{equation}
L_{0{\rm gc}}^{-1}F \;\equiv\; \Omega^{-1}\;\int \wt{F}\, d\theta \;\equiv\; \Omega^{-1}\;\int \left( F \;-\frac{}{} \langle F\rangle \right)d\theta,
\label{eq:L0gc_inv}
\end{equation}
where $\wt{F}$ denotes the gyroangle-dependent part of $F$ and $\langle F\rangle$ denotes the gyroangle-averaged part. We immediately see that the zeroth-order operator explicitly involves the lowest-order guiding-center transformation (for a uniform magnetic field) and we note that this formulation can also be applied to the bounce-motion Vlasov description of magnetically-trapped particles \cite{Brizard_Hahm}.

\subsection{First-Order Vlasov Operator}

The first-order Vlasov operator
\begin{equation}
L_{1} \;=\; \left( \pd{}{t} \;+\; v_{\|}\,\bhat\bdot\nabla \;+\; \frac{q}{m}\,E_{\|}\;\pd{}{v_{\|}} \right) \;+\; \left( \frac{q}{m}\,
{\bf E}_{\bot}\bdot\pd{}{{\bf v}_{\bot}} \right) \;\equiv\; L_{1\|} \;+\; L_{1\bot}
\label{eq:L1_def}
\end{equation}
is explicitly decomposed in terms of the parallel and perpendicular components of the electric field. The ordering $L_{0} + \epsilon\,L_{1}$ in
(\ref{eq:Vlasov_L}) implies that the gyromotion time scale is the shortest time scale in our plasma physics problem. In particular, the ordering 
$\Omega^{-1}L_{1\bot} \ll 1$ implies that the perpendicular $E\times B$ velocity is small compared to the characteristic (i.e., thermal) velocity of a particle (this is the drift ordering \cite{HH}). The ordering $\Omega^{-1}L_{1\|} \ll 1$, on the other hand, implies that the time scale of interest is long compared to the gyration period and that the parallel gradient length scale is long compared to the gyroradius. These two orderings are consistent with the guiding-center and gyrocenter orderings \cite{Brizard_Hahm}. [Note that it is also possible to work with an alternate ordering where $\Omega^{-1}L_{1\|} \ll \Omega^{-1}L_{1\bot}$, in which case one could elevate $L_{1\|} \equiv L_{2}$ to a higher order in (\ref{eq:Vlasov_hierarchy}).]

We first look at the operator $L_{1\bot}$ in (\ref{eq:L1_def}), which can be rewritten as
\begin{eqnarray} 
L_{1\bot}G & \equiv & q\,\frac{\Omega}{B}\;{\bf E}_{\bot}\bdot \left( \pd{\vb{\rho}_{\rm gc}}{\theta}\;
\pd{G}{\mu} \;-\; \pd{\vb{\rho}_{\rm gc}}{\mu}\;\pd{G}{\theta} \right) \nonumber \\
 & = & -\;q\,\frac{\Omega}{B} \left[  \left( \nabla\phi\bdot \pd{\vb{\rho}_{\rm gc}}{\theta} \right)
\pd{G}{\mu} \;-\; \left( \nabla\phi\bdot\pd{\vb{\rho}_{\rm gc}}{\mu} \right) \pd{G}{\theta} \right],
\label{eq:L1_bot}
\end{eqnarray}
where $\mu \equiv m|{\bf v}_{\bot}|^{2}/2B$ denotes the guiding-center magnetic moment (a guiding-center invariant in a uniform magnetic field) and 
$\vb{\rho}_{\rm gc}$ denotes the gyroradius vector. We now use the identity
\[ \pd{\vb{\rho}_{\rm gc}}{J^{i}}\bdot\nabla\phi \;\equiv\; \left( {\sf T}_{\rm gc}\;\pd{}{J^{i}}\;{\sf T}_{\rm gc}^{-1}\right)\phi \;=\; {\sf T}_{{\rm gc}}\left( 
\pd{\phi_{{\rm gc}}}{J^{i}} \right), \]
where $\partial/\partial J^{i} \equiv (\partial/\partial\theta,\; \partial/\partial\mu)$ and $\phi_{{\rm gc}} \equiv {\sf T}_{{\rm gc}}^{-1}\phi = 
\phi({\bf X} + \vb{\rho}_{\rm gc})$ defines the {\it guiding-center} scalar potential, so that the operator (\ref{eq:L1_bot}) can be written as
\[ L_{1\bot}G \;=\; -\;q\,\frac{\Omega}{B} \left[ \left( {\sf T}_{{\rm gc}} \pd{\phi_{{\rm gc}}}{\theta}\right)\pd{}{\mu} 
\;-\; \left( {\sf T}_{{\rm gc}} \pd{\phi_{{\rm gc}}}{\mu}\right)\pd{}{\theta} \right] {\sf T}_{{\rm gc}}G_{{\rm gc}}, \]
where $G \equiv {\sf T}_{{\rm gc}}G_{{\rm gc}}$ is expressed in terms of the pull-back of $G_{{\rm gc}}$. Next, we note that the operators $\partial/\partial J^{i}$ and ${\sf T}_{{\rm gc}} \equiv \exp(-\,\vb{\rho}_{\rm gc}\bdot\nabla)$ do not commute:
\[ \left[ \pd{}{J^{i}},\; {\sf T}_{{\rm gc}}\right]G_{{\rm gc}} \;\equiv\; \pd{}{J^{i}}\left({\sf T}_{{\rm gc}}G_{{\rm gc}}\frac{}{}\right) \;-\; 
{\sf T}_{{\rm gc}} \left( \pd{G_{{\rm gc}}}{J^{i}} \right) \;=\; -\;{\sf T}_{{\rm gc}} \left( \pd{\vb{\rho}_{\rm gc}}{J^{i}}\bdot\nabla G_{{\rm gc}} \right). \]
Hence, using the fact that the pull-back operator is distributive ${\sf T}_{{\rm gc}}(F\;G) \equiv ({\sf T}_{{\rm gc}}F)\;({\sf T}_{{\rm gc}}G)$, 
we obtain
\begin{eqnarray}
L_{1\bot}G & \equiv & -\;q\,\frac{\Omega}{B}\;{\sf T}_{{\rm gc}} \left[\; \left( \pd{\phi_{{\rm gc}}}{\theta}\;\pd{G_{{\rm gc}}}{\mu} \;-\; 
\pd{\phi_{{\rm gc}}}{\mu}\;\pd{G_{{\rm gc}}}{\theta} \right) \right. \nonumber \\
 &  &\left.\hspace*{0.8in}-\; \nabla\phi_{{\rm gc}}\bdot\left( \pd{\vb{\rho}_{\rm gc}}{\theta}\;\pd{\vb{\rho}_{\rm gc}}{\mu} \;-\; 
\pd{\vb{\rho}_{\rm gc}}{\mu}\;\pd{\vb{\rho}_{\rm gc}}{\theta} \right)\bdot\nabla G_{{\rm gc}} \;\right],
\label{eq:L1bot_F}
\end{eqnarray}
where we have used the explicit property $\partial\phi_{{\rm gc}}/\partial J^{i} = \nabla\phi_{{\rm gc}}\bdot\partial\vb{\rho}_{\rm gc}/\partial J^{i}$ to obtain the last two terms. Lastly, by using the identity (valid for any vectors ${\bf F}$ and ${\bf G}$)
\[ {\bf F}\bdot\left(\pd{\vb{\rho}_{\rm gc}}{\theta}\;\pd{\vb{\rho}_{\rm gc}}{\mu} \;-\; \pd{\vb{\rho}_{\rm gc}}{\mu}\;\pd{\vb{\rho}_{\rm gc}}{\theta}\right)\bdot{\bf G} \;\equiv\;
\frac{{\bf B}}{m\Omega^{2}}\bdot({\bf F}\btimes{\bf G}), \]
we obtain our final expression for $L_{1\bot}$:
\begin{eqnarray}
L_{1\bot}G & = & -\,q\;{\sf T}_{{\rm gc}}\left[ \frac{\Omega}{B} \left( \pd{\phi_{{\rm gc}}}{\theta}\;\pd{G_{{\rm gc}}}{\mu} \;-\; 
\pd{\phi_{{\rm gc}}}{\mu}\;\pd{G_{{\rm gc}}}{\theta} \right) \;-\; \frac{\bhat}{m\Omega}\bdot\left( \nabla\phi_{{\rm gc}}\btimes\nabla G_{{\rm gc}}
\frac{}{}\right) \;\right] \nonumber \\
 & \equiv & -\;q\;{\sf T}_{{\rm gc}}\left( \left\{ \phi_{{\rm gc}},\; G_{{\rm gc}} \right\}_{\bot{\rm gc}}\right),
\label{eq:L1bot_gc}
\end{eqnarray}
where we introduced the ``perpendicular'' components of the guiding-center Poisson bracket. 

We can similarly write $L_{1\|}G$ in terms of $\phi_{{\rm gc}}$ and $G_{{\rm gc}}$ as
\begin{equation}
L_{1\|}G \;\equiv\; {\sf T}_{{\rm gc}} \left( \frac{d_{{\rm gc}}}{dt}\,G_{{\rm gc}} \;+\; q \left\{ G_{{\rm gc}},\; \phi_{{\rm gc}} \right\}_{\|{\rm gc}}
\right),
\label{eq:L1par_gc}
\end{equation}
where $d_{{\rm gc}}/dt \equiv \partial/\partial t + v_{\|}\,\bhat\bdot\nabla$ is the guiding-center Vlasov operator (in a uniform magnetic field), and
\[ \left\{ F,\; G \right\}_{\|{\rm gc}} \;\equiv\; \frac{\bhat}{m}\bdot\left( \nabla F\;\pd{G}{v_{\|}} \;-\; \pd{F}{v_{\|}}\;\nabla G \right) \]
denotes the ``parallel'' guiding-center Poisson bracket. By combining the perpendicular and parallel components (\ref{eq:L1bot_gc}) and 
(\ref{eq:L1par_gc}), we therefore obtain the final expression for the first-order Vlasov operator
\begin{equation}
L_{1}G \;=\; {\sf T}_{{\rm gc}} \left( \frac{d_{{\rm gc}}}{dt}\,G_{{\rm gc}} \;+\; q\;\left\{ G_{{\rm gc}},\; \phi_{{\rm gc}} \right\}_{{\rm gc}}
\right) \;\equiv\; {\sf T}_{{\rm gc}} \left(\,L_{1{\rm gc}}\,G_{{\rm gc}} \frac{}{}\right),
\label{eq:L1_gc}
\end{equation}
where $\{\;,\;\}_{{\rm gc}}$ now denotes the full guiding-center Poisson bracket. We note that the guiding-center representation (\ref{eq:L1_gc}) also holds for a nonuniform background magnetic field, where the guiding-center Poisson bracket now contains corrections associated with magnetic-field nonuniformity (see Appendix A).

Lastly we note that it was the original insight of Catto \cite{Catto} to recognize that the transformation from particle coordinates to guiding-center coordinates greatly simplifies the recursive solution of the Vlasov equation (\ref{eq:Vlasov_eq}) within the context of linear gyrokinetic theory. What we have shown in this Section is that this simplification naturally extends to the inclusion of the guiding-center Poisson bracket $\{\;,\;\}_{{\rm gc}}$ as well as finite-Larmor-radius (FLR) effects associated with the scalar potential $\phi$ (through the pull-back and push-forward operators) in the first-order guiding-center Vlasov operator $L_{1{\rm gc}}$.

\section{\label{sec:gcrv}Guiding-center recursive Vlasov method}

In this Section, we use the guiding-center recursive Vlasov (gcrV) method to derive asymptotic expansions for the gyrocenter phase-space coordinates. Here, the gcrV method is defined in terms of recursive solutions involving the guiding-center Vlasov operator
\begin{equation}
L_{{\rm gc}} \;\equiv\; {\sf T}_{{\rm gc}}^{-1}\,L\;{\sf T}_{{\rm gc}} \;=\; \Omega\;\pd{}{\theta} \;+\; \epsilon \left( \frac{d_{{\rm gc}}}{dt} \;+\; q\;\left\{ \;\;,\; \phi_{{\rm gc}} \right\}_{{\rm gc}} \right) \;=\; L_{0{\rm gc}} \;+\; \epsilon\;L_{1{\rm gc}}.
\label{eq:L_gc}
\end{equation}
The gcrV method therefore naturally takes into account the full FLR effects of the perturbation scalar potential.

A generic gyrocenter variable $Q_{\rm gy}$ can be expressed in terms of a time-dependent asymptotic expansion 
\begin{equation}
Q_{{\rm gy}} \;=\; Q^{(0)}_{\rm gy} \;+\; \epsilon\,Q^{(1)}_{\rm gy} \;+\; \epsilon^{2}\,Q^{(2)}_{\rm gy} \;+\; \cdots \;\equiv\; 
{\sf T}_{\rm gy}^{-1}Q_{\rm gc} \;=\; {\sf T}_{\rm gy}^{-1}\left({\sf T}_{\rm gc}^{-1}Q\right), 
\label{eq:zalpha_asym}
\end{equation}
where $Q^{(0)}_{\rm gy} \equiv Q_{\rm gc}$ denotes the 
guiding-center variable and the $k$th-order gyrocenter variable $Q^{(k)}_{\rm gy} \equiv \ov{Q}^{(k)}_{\rm gy} + \wt{Q}^{(k)}_{\rm gy}$ is decomposed in terms of gyroangle-independent and gyroangle-dependent parts (respectively). The asymptotic expansion
(\ref{eq:zalpha_asym}) is constructed by gcrV method by requiring that the quantity $\dot{Q}_{\rm gy} =
L_{\rm gc} Q_{\rm gy}$ is gyrophase independent. This condition yields the following $k$th-order expressions
\begin{equation}
\dot{Q}_{\rm gy}^{(k)} \;\equiv\; \left\langle L_{1 \rm gc} Q^{(k-1)}_{\rm gy}\right\rangle \;=\; \langle L_{1 \rm gc}\rangle\,\ov{Q}^{(k-1)}_{\rm gy} \;+\;\left\langle \wt{L}_{1{\rm gc}}\,\wt{Q}^{(k-1)}_{\rm gy}\right\rangle,
\label{eq:Qk_dot}
\end{equation}
and
\be
\wt{Q}^{(k)}_{\rm gy} \;=\; -\;L_{0{\rm gc}}^{-1}\left[ L_{1 \rm gc} Q^{(k-1)}_{\rm gy} \right].
\label{eq:gyro_dep}
\ee
Note that the solution for the gyroangle-independent part $\ov{Q}^{(k-1)}_{\rm gy}$ appears at the $k$th order.

The gcrV method now proceeds as follows. At zeroth-order, for the gyrocenter phase-space variable $Z_{0{\rm gy}}^{\alpha} \neq \theta$, we find the definition
\begin{equation}
\left( \dot{Z}_{{\rm gy}}^{\alpha}\right)_{0} \;\equiv\; \Omega\;\left\langle \pd{Z_{0{\rm gy}}^{\alpha}}{\theta} \right\rangle \;=\; 0,
\label{eq:ivdgy_0}
\end{equation}
so that the zeroth-order gyrocenter (guiding-center) coordinate $Z_{0{\rm gy}}^{\alpha} \neq \theta$ must be independent of the gyroangle $\theta$ (i.e., $\langle Z_{0{\rm gy}}^{\alpha}\rangle \equiv Z_{0{\rm gy}}^{\alpha}$). Obviously, $(\dot{\theta}_{{\rm gy}})_{0} \equiv \Omega$ for the gyrocenter gyroangle $Z_{0{\rm gy}}^{\alpha} = \theta$. 

At first order, we find the gyroangle-independent expression
\begin{equation}
\left( \dot{Z}_{{\rm gy}}^{\alpha}\right)_{1} \;\equiv\; \left\langle L_{1{\rm gc}}\right\rangle\,Z_{0{\rm gy}}^{\alpha} \;=\; 
\frac{d_{{\rm gc}}Z_{0{\rm gy}}^{\alpha}}{dt} \;+\; q\;\left\{ Z_{0{\rm gy}}^{\alpha},\; \langle\phi_{{\rm gc}}\rangle\right\}_{{\rm gc}} \;\equiv\; \frac{d_{{\rm gy}}Z_{0{\rm gy}}^{\alpha}}{dt},
\label{eq:ivdgy_1}
\end{equation}
where $d_{{\rm gy}}/dt$ denotes the lowest-order gyrocenter Vlasov operator (which includes the $E\times B$ velocity and parallel electric field associated with $\langle\phi_{{\rm gc}}\rangle$). The gyroangle-dependent first-order expression, on the other hand, is obtained from (\ref{eq:gyro_dep}) as
\begin{equation}
\wt{Z}_{1{\rm gy}}^{\alpha} \;\equiv\; -\;L_{0{\rm gc}}^{-1} \left( q\,\left\{ Z_{0{\rm gy}}^{\alpha},\frac{}{} \phi_{\rm gc} \right\}_{\rm gc} \right) 
\;=\; -\;\frac{q}{\Omega}\;\left\{ Z_{0{\rm gy}}^{\alpha},\; \wt{\Phi}_{{\rm gc}} \right\}_{{\rm gc}},
\label{eq:ivdgc_1}
\end{equation}
where 
\begin{equation}
\wt{\Phi}_{{\rm gc}} \;\equiv\; \int\,\wt{\phi}_{{\rm gc}}\,d\theta \;=\; \int \left( \phi_{{\rm gc}} \;-\; \langle\phi_{{\rm gc}}\rangle \frac{}{}
\right) d\theta 
\label{eq:Phi_gc}
\end{equation}
denotes the indefinite gyroangle integral of the gyroangle-dependent part of $\phi_{{\rm gc}}$. Note that the gyroangle-independent part of 
$Z_{1{\rm gy}}^{\alpha}$ (denoted $\ov{Z}_{1{\rm gy}}^{\alpha} \equiv Z_{1{\rm gy}}^{\alpha} - \wt{Z}_{1{\rm gy}}^{\alpha}$) must be determined at the second order.

At second order, we find the gyroangle-independent expression
\begin{equation}
\left( \dot{Z}_{{\rm gy}}^{\alpha}\right)_{2} \;\equiv\; \frac{d_{{\rm gy}}\ov{Z}_{1{\rm gy}}^{\alpha}}{dt} \;+\; q\;\left\langle \left\{ 
\wt{Z}_{1{\rm gy}}^{\alpha},\; \wt{\phi}_{{\rm gc}} \right\}_{{\rm gc}} \right\rangle,
\label{eq:ivdgy_2}
\end{equation}
and the gyroangle-dependent expression
\begin{equation}
\wt{Z}_{2{\rm gy}}^{\alpha} \;=\; -\;\frac{q}{\Omega}\left\{ \ov{Z}_{1{\rm gy}}^{\alpha},\; \wt{\Phi}_{{\rm gc}} \right\}_{{\rm gc}} \;-\;
L_{0{\rm gc}}^{-1}\left( L_{1{\rm gc}}\,\wt{Z}_{1{\rm gy}}^{\alpha}\right). 
\label{eq:ivdgc_2}
\end{equation}
In (\ref{eq:ivdgy_2}), we see that the role of $\ov{Z}_{1{\rm gy}}^{\alpha}$ is to ensure that $(\dot{Z}_{{\rm gy}}^{\alpha})_{2}$ satisfies any desired property we want (e.g., be gyroangle-independent or zero). One clearly sees how the gcrV method can be extended to higher order.

\subsection{Gyrocenter magnetic moment}

The easiest gyrocenter phase-space variable one can construct by the gcrV method is the gyrocenter magnetic moment 
$\mu_{{\rm gy}}$ since it is to be constructed as an invariant of the gyrocenter dynamics, i.e., $\dot{\mu}_{{\rm gy}} \equiv 0$ to arbitrary order in 
$\epsilon$. At zeroth order, we easily find $L_{0{\rm gc}}\,\mu \equiv 0$, where $\mu_{0{\rm gy}} \equiv \mu$ denotes the guiding-center magnetic moment (which satisfies $d_{{\rm gc}}\mu/dt \equiv 0$). 

At first order, the requirement $(\dot{\mu}_{{\rm gy}})_{1} \equiv 0$ becomes
\begin{equation}
0 \;=\; \frac{d_{{\rm gc}}\mu}{dt} \;+\; q\;\left\{ \mu,\; \phi_{{\rm gc}} \right\}_{{\rm gc}} \;+\;
\Omega\;\pd{\mu_{1{\rm gy}}}{\theta} \;=\; \Omega\;\pd{}{\theta} \left( \mu_{1{\rm gy}} \;-\; \frac{q}{B}\;\phi_{{\rm gc}} \right),
\label{eq:mu1_dot}
\end{equation}
which is easily solved as
\begin{equation}
\mu_{1{\rm gy}} \;\equiv\; \ov{\mu}_{1{\rm gy}} \;+\; \frac{q}{B}\;\wt{\phi}_{{\rm gc}} \;=\; \left( \ov{\mu}_{1{\rm gy}} \;-\; \frac{q}{B}\;
\langle\phi_{{\rm gc}}\rangle \right) \;+\; \frac{q}{B}\;\phi_{{\rm gc}},
\label{eq:mu1_gy}
\end{equation}
where $\ov{\mu}_{1{\rm gy}}$ denotes the gyroangle-independent part of the first-order gyrocenter magnetic moment (to be determined at the second order). We note that the first-order magnetic-moment correction $\Delta\mu$ derived by Taylor \cite{Taylor} can be expressed as the guiding-center pull-back 
$\Delta\mu \equiv {\sf T}_{\rm gc}\;\wt{\mu}_{1{\rm gy}}$ of the gyrocenter magnetic-moment correction (\ref{eq:mu1_gy}), where
\begin{equation}
{\sf T}_{{\rm gc}}\wt{\phi}_{{\rm gc}} \;=\; {\sf T}_{\rm gc}\left( {\sf T}_{\rm gc}^{-1}\phi \;-\; \langle\phi_{\rm gc}\rangle\right) \;\equiv\; \phi \;-\; {\sf T}_{{\rm gc}}\langle\phi_{{\rm gc}}\rangle.
\label{eq:Tphi_gc}
\end{equation}

At second order, the requirement $(\dot{\mu}_{{\rm gy}})_{2} \equiv 0$ becomes
\begin{eqnarray}
0 & = & \frac{d_{{\rm gc}}\mu_{1{\rm gy}}}{dt} \;+\; q\;\left\{ \mu_{1{\rm gy}},\; \phi_{{\rm gc}} \right\}_{{\rm gc}} \;+\; \Omega\;\pd{\mu_{2{\rm gy}}}{\theta} \label{eq:mu2_dot} \\
 & = & \frac{d_{{\rm gc}}}{dt} \left( \ov{\mu}_{1{\rm gy}} \;+\; \wt{\mu}_{1{\rm gy}} \right) \;+\; q\;\left\{ 
\left( \ov{\mu}_{1{\rm gy}} \;-\; \frac{q}{B}\,\langle\phi_{{\rm gc}}\rangle \right),\; \phi_{{\rm gc}} \right\}_{{\rm gc}} \;+\; 
\Omega\;\pd{\mu_{2{\rm gy}}}{\theta},
\nonumber
\end{eqnarray}
where we used (\ref{eq:mu1_gy}) for $\mu_{1{\rm gy}}$ (with $\{ \phi_{{\rm gc}},\; \phi_{{\rm gc}}\}_{{\rm gc}} \equiv 0$) and the role of 
$\ov{\mu}_{1{\rm gy}}$ is to ensure that the gyroangle-independent right side of (\ref{eq:mu2_dot}) is zero. This condition yields the homogeneous equation $d_{{\rm gy}}\ov{\mu}_{1{\rm gy}}/dt = 0$, whose solution is simply $\ov{\mu}_{1{\rm gy}} \equiv 0$. The solution to the gyroangle-dependent part of (\ref{eq:mu2_dot}) is finally expressed as
\begin{equation}
\mu_{2{\rm gy}} \;\equiv\; \ov{\mu}_{2{\rm gy}} \;-\; \frac{q}{B\Omega}\;\frac{d_{{\rm gy}}\wt{\Phi}_{{\rm gc}}}{dt},
\label{eq:mu2_gy}
\end{equation}
where 
\[ \frac{d_{{\rm gy}}\wt{\Phi}_{{\rm gc}}}{dt} \;=\; \left( \frac{d_{{\rm gc}}}{dt} \;+\; \frac{c\bhat}{B}\btimes\nabla\langle\phi_{\rm gc}\rangle
\bdot\nabla\right) \wt{\Phi}_{{\rm gc}} \;+\; \Omega \left( \frac{q}{B}\,\pd{\langle\phi_{\rm gc}\rangle}{\mu}\right) \wt{\phi}_{\rm gc}, \]
and the gyroangle-independent part $\ov{\mu}_{2{\rm gy}}$ is determined at the third order.

Lastly, it is instructive to compute the third-order component of the gyrocenter magnetic moment. Here, $(\dot{\mu}_{{\rm gy}})_{3} \equiv 0$ becomes
\begin{eqnarray}
0 & = & \frac{d_{{\rm gc}}\mu_{2{\rm gy}}}{dt} \;+\; q\;\left\{ \mu_{2{\rm gy}},\; 
\phi_{{\rm gc}}  \right\}_{{\rm gc}} \;+\; \Omega\;\pd{\mu_{3{\rm gy}}}{\theta} \nonumber \\
 & = & \frac{d_{{\rm gy}}}{dt} \left( \ov{\mu}_{2{\rm gy}} \;+\; \wt{\mu}_{2{\rm gy}} \right) \;+\; q\;\left\{ \left( \ov{\mu}_{2{\rm gy}} \;+\; 
\wt{\mu}_{2{\rm gy}} \right),\; \wt{\phi}_{{\rm gc}} \right\}_{{\rm gc}} \;+\; \Omega\;\pd{\mu_{3{\rm gy}}}{\theta}.
\label{eq:mu3_dot}
\end{eqnarray}
The role of $\ov{\mu}_{2{\rm gy}}$ is to ensure that the gyroangle-independent right side of (\ref{eq:mu3_dot}) is zero, which implies that 
$\ov{\mu}_{2{\rm gy}}$ is a solution of the inhomogeneous equation
\[ \frac{d_{{\rm gy}}\ov{\mu}_{2{\rm gy}}}{dt} \;=\; q\;\left\langle \left\{ \wt{\phi}_{{\rm gc}},\; \wt{\mu}_{2{\rm gy}}\right\}_{{\rm gc}} 
\right\rangle \;=\; -\; \frac{q^{2}}{B\Omega} \left\langle \left\{ \wt{\phi}_{{\rm gc}},\; \frac{d_{{\rm gy}}\wt{\Phi}_{{\rm gc}}}{dt}
\right\}_{{\rm gc}} \right\rangle. \]
Because the right side of this equation is non-vanishing, there must be a non-trivial solution for $\ov{\mu}_{2{\rm gy}}$. By using properties of the guiding-center Poisson bracket $\{\;,\;\}_{{\rm gc}}$ and the gyrocenter Vlasov operator $d_{{\rm gy}}/dt$ (after tedious algebra), we find the 
second-order solution 
\begin{equation}
\ov{\mu}_{2{\rm gy}} \;\equiv\; \frac{q^{2}}{2B\Omega}\;\left\langle \left\{ \wt{\Phi}_{{\rm gc}},\; \wt{\phi}_{{\rm gc}} \right\}_{{\rm gc}} \right\rangle.
\label{eq:ovmu2_gy}
\end{equation}
This solution is more trivially obtained with the Lie-transform approach presented in \S\,\ref{sec:gyro_Lie}. To lowest FLR order, where $\wt{\phi}_{{\rm gc}} \equiv \partial\wt{\Phi}_{{\rm gc}}/\partial\theta \simeq \vb{\rho}_{\rm gc}\bdot\nabla_{\bot}\phi$, we note that
\[ \left\langle \left\{ \wt{\Phi}_{{\rm gc}},\; \wt{\phi}_{{\rm gc}} \right\}_{{\rm gc}} \right\rangle \;=\; \frac{\Omega}{B}\,
\pd{\langle(\wt{\phi}_{{\rm gc}})^{2}\rangle}{\mu} \;+\; \cdots \;=\; \frac{|\nabla_{\bot}\phi|^{2}}{m\,\Omega}, \]
so that
\begin{equation}
\ov{\mu}_{2{\rm gy}} \;\simeq\; \frac{m}{2B}\;|{\bf u}_{{\rm E}}|^{2},
\label{eq:ovmu2_ExB}
\end{equation}
where ${\bf u}_{{\rm E}} \equiv (c\bhat/B)\btimes\nabla_{\bot}\phi$ denotes the $E\times B$ velocity. We omit the explicit derivation of $\wt{\mu}_{3{\rm gy}}$, which is obtained from the gyroangle-dependent of the third-order equation (\ref{eq:mu3_dot}) and contains terms that are of second order in $\Omega^{-1}$.

Up to second order in $\epsilon$ (and first order in $\Omega^{-1}$), the gyrocenter magnetic moment is therefore expressed as
\begin{equation}
\mu_{{\rm gy}} \;=\; \mu \;+\; \frac{q}{B} \left( \wt{\phi}_{{\rm gc}} \;-\; \frac{1}{\Omega}\;\frac{d_{{\rm gy}}\wt{\Phi}_{{\rm gc}}}{dt} 
\right) \;+\; \frac{q^{2}}{2B\Omega}\;\left\langle \left\{ \wt{\Phi}_{{\rm gc}},\; \wt{\phi}_{{\rm gc}} \right\}_{{\rm gc}} \right\rangle.
\label{eq:mu_gy}
\end{equation}
We note that, to lowest FLR order, the first two terms in (\ref{eq:mu_gy}) appear naturally in the expansion of the magnetic moment
$\mu_{\rm gy} = m\,|{\bf v}_{\bot} - ({\bf u}_{E} + {\bf u}_{P})|^{2}/2B$, where ${\bf u}_{P} = -\,(c/B\Omega)\,d\nabla\phi/dt$ denotes the polarization drift velocity. Lastly, we note that Parra \& Catto \cite{P_Catto} only computed the first-order correction $(q\,\wt{\phi}_{{\rm gc}}/B)$ to the gyrocenter magnetic moment. While the polarization-drift correction $d_{{\rm gy}}\wt{\Phi}_{{\rm gc}}/dt$ is generally not kept in standard gyrokinetic theory \cite{Brizard_Hahm}, we show in \S\,\ref{sec:gyro_Lie} how it also appears naturally in the Lie-transform approach. 

\subsection{Gyrocenter gyroangle}

At zeroth order, we easily find $L_{0{\rm gc}}\theta \equiv \Omega$, where $\theta_{0{\rm gy}} \equiv \theta$ denotes the guiding-center gyroangle. At first order, we find
\begin{equation}
\left( \dot{\theta}_{{\rm gy}}\right)_{1} \;=\; \Omega \left( \frac{q}{B}\;\pd{\phi_{{\rm gc}}}{\mu} \;+\; \pd{\theta_{1{\rm gy}}}{\theta} \right),
\label{eq:theta1_dot}
\end{equation}
which yields the gyroangle-independent equation
\begin{equation}
\left( \dot{\theta}_{{\rm gy}}\right)_{1} \;=\; \frac{q\Omega}{B}\;\pd{\langle\phi_{{\rm gc}}\rangle}{\mu},
\label{eq:theta1gy_dot}
\end{equation}
while the gyroangle-dependent equation yields the solution
\begin{equation}
\wt{\theta}_{1{\rm gy}} \;=\; -\; \frac{q}{B}\;\pd{\wt{\Phi}_{{\rm gc}}}{\mu} \;\equiv\; \theta_{1{\rm gy}} \;-\; \ov{\theta}_{1{\rm gy}},
\label{eq:theta1_gy}
\end{equation}
where $\ov{\theta}_{1{\rm gy}}$ denotes the gyroangle-independent part to be determined at second order.

\subsection{Gyrocenter position}

The zeroth-order gyrocenter position is the guiding-center position ${\bf X}_{0{\rm gy}} = {\bf X} \equiv {\bf r} - \vb{\rho}_{\rm gc}$, where ${\bf r}$ is the particle's position and $\vb{\rho}_{\rm gc}$ is the gyroangle-dependent gyroradius vector. At zeroth-order, we easily find $L_{0{\rm gc}}{\bf X} \equiv 0$, since the guiding-center position ${\bf X}$ is independent of $\theta$. At first order, we find
\begin{eqnarray}
\left( \dot{{\bf X}}_{{\rm gy}}\right)_{1} & = & \frac{d_{{\rm gc}}{\bf X}}{dt} \;+\; q\;\left\{ {\bf X},\; \phi_{{\rm gc}} 
\right\}_{{\rm gc}} \;+\; \Omega\;\pd{{\bf X}_{1{\rm gy}}}{\theta} \nonumber \\
 & = & v_{\|}\,\bhat \;+\; \frac{c\bhat}{B}\btimes\nabla\phi_{{\rm gc}} \;+\; \Omega\;
\pd{{\bf X}_{1{\rm gy}}}{\theta},
\label{eq:R1_dot}
\end{eqnarray}
which yields the gyroangle-independent first-order equation for the gyrocenter velocity
\begin{equation}
\left( \dot{{\bf X}}_{{\rm gy}}\right)_{1} \;=\; v_{\|}\,\bhat \;+\; \frac{c\bhat}{B}\btimes\nabla\langle\phi_{{\rm gc}}\rangle.
\label{eq:R1gy_dot}
\end{equation}
We therefore see that the lowest-order gyrocenter motion is described in terms of parallel motion along and $E\times B$ motion across the field lines.
The gyroangle-dependent equation obtained from (\ref{eq:R1_dot}) yields the solution
\begin{equation}
\wt{{\bf X}}_{1{\rm gy}} \;=\; -\; \frac{c\bhat}{B\Omega}\btimes\nabla\wt{\Phi}_{{\rm gc}} \;\equiv\; {\bf X}_{1{\rm gy}} \;-\; 
\ov{{\bf X}}_{1{\rm gy}},
\label{eq:R1_gy}
\end{equation}
where $\ov{{\bf X}}_{1{\rm gy}}$ denotes the gyroangle-independent part to be determined at second order.

\subsection{Gyrocenter parallel momentum}

The zeroth-order gyrocenter parallel momentum is the guiding-center parallel momentum $p_{\|0{\rm gy}} \equiv mv_{\|}$. At first order, we find 
\begin{equation}
\left(\dot{p}_{\|{\rm gy}}\right)_{1} \;=\; -\;q\,\bhat\bdot\nabla\phi_{{\rm gc}} \;+\; \Omega\;\pd{p_{\|1{\rm gy}}}{\theta}.
\label{eq:vpargy_1}
\end{equation}
The gyroangle-independent part of this equation yields
\begin{equation}
\left(\dot{p}_{\|{\rm gy}}\right)_{1} \;=\; -\;q\,\bhat\bdot\nabla\langle\phi_{{\rm gc}}\rangle,
\label{eq:vpargy_dot}
\end{equation}
while the gyroangle-dependent part yields
\begin{equation}
\wt{p}_{\|1{\rm gy}} \;=\; q\;\frac{\bhat}{\Omega}\bdot\nabla\wt{\Phi}_{{\rm gc}} \;\equiv\; p_{\|1{\rm gy}} \;-\; 
\ov{p}_{\|1{\rm gy}},
\label{eq:vpar_gy}
\end{equation}
where $\ov{p}_{\|1{\rm gy}}$ denotes the gyroangle-independent part to be determined at second order.

\subsection{\label{subsec:kin_energy_it}Gyrocenter kinetic energy}

We now use the operators $L_{0{\rm gc}}$ and $L_{1{\rm gc}}$ to derive an asymptotic expansion for the kinetic energy $K_{\rm gy} = K_{0{\rm gy}} + \epsilon\,K_{1{\rm gy}} + \cdots$, where 
\[ K_{0{\rm gy}} \;=\; mv^{2}/2 \;\equiv\; p_{\|}^{2}/2m \;+\; \mu\,B \;=\; K_{\rm gc} \]
is expressed in terms of the lowest-order guiding-center coordinates $p_{\|}$ and $\mu$. We want to construct $K_{\rm gy}$ such that $\dot{K}_{\rm gy}$ is gyroangle independent. At zeroth order, we easily find that $(\dot{K}_{{\rm gy}})_{0} = L_{0{\rm gc}}K_{0{\rm gy}} \equiv 0$, i.e., the guiding-center kinetic energy $K_{\rm gc}$ is a constant on the gyromotion time scale. 

At first order, we find 
\begin{eqnarray}
(\dot{K}_{{\rm gy}})_{1} & = & \frac{d_{{\rm gc}}K_{0{\rm gy}}}{dt} \;+\; q\;\left\{ K_{0{\rm gy}},\; \phi_{{\rm gc}} 
\right\}_{{\rm gc}} \;+\; \Omega\,\pd{K_{1{\rm gy}}}{\theta} \nonumber \\
 & = & -\;q \left( \Omega\;\pd{\phi_{{\rm gc}}}{\theta} \;+\; v_{\|}\;
\bhat\bdot\nabla\phi_{{\rm gc}} \right) \;+\; \Omega\,\pd{K_{1{\rm gy}}}{\theta},
\label{eq:K1_eq}
\end{eqnarray}
where we used the fact that $d_{{\rm gc}}K_{0{\rm gy}}/dt \equiv 0$. The gyroangle-independent and gyroangle-dependent parts of (\ref{eq:K1_eq}) yield 
\begin{equation}
(\dot{K}_{{\rm gy}})_{1} \;=\; -\;q\,v_{\|}\;\bhat\bdot\nabla\langle\phi_{{\rm gc}}\rangle, 
\label{eq:Kgy1_dot}
\end{equation}
and 
\begin{equation} 
\wt{K}_{1{\rm gy}} \;=\; q\;\wt{\phi}_{{\rm gc}} \;+\; \frac{q}{\Omega}\,v_{\|}\;\bhat\bdot\nabla\wt{\Phi}_{{\rm gc}} \;=\; 
\frac{q}{\Omega}\;\left\{ \wt{\Phi}_{{\rm gc}},\; K_{\rm gc} \right\}_{{\rm gc}} \;\equiv\; K_{1{\rm gy}} \;-\; \ov{K}_{1{\rm gy}},
\label{eq:K1_gc}
\end{equation}
where the gyroangle-independent part $\ov{K}_{1{\rm gy}}$ is determined at the next order.

At second order, we find 
\begin{eqnarray}
(\dot{K}_{{\rm gy}})_{2} & = &  \frac{d_{{\rm gc}}K_{1{\rm gy}}}{dt} \;+\; q\;\left\{ K_{1{\rm gy}},\; \phi_{{\rm gc}}\right\}_{{\rm gc}} \;+\; \Omega\,\pd{K_{2{\rm gy}}}{\theta} \label{eq:K2_eq} \\
 & = & \frac{d_{{\rm gy}}}{dt}\left( \ov{K}_{1{\rm gy}} \;+\; \wt{K}_{1{\rm gy}} \right) \;+\; q\;\left\{ \left( \ov{K}_{1{\rm gy}} \;+\; 
\wt{K}_{1{\rm gy}} \right),\; \wt{\phi}_{{\rm gc}} \right\}_{{\rm gc}} \;+\; \Omega\,\pd{K_{2{\rm gy}}}{\theta}. \nonumber 
\end{eqnarray}
By choosing $\ov{K}_{1{\rm gy}} \equiv 0$ in (\ref{eq:K2_eq}), we obtain the gyroangle-independent part
\begin{equation}
(\dot{K}_{{\rm gy}})_{2} \;\equiv\; q\;\left\langle\left\{ \wt{K}_{1{\rm gy}},\; \wt{\phi}_{{\rm gc}} \right\}_{{\rm gc}}\right\rangle.
\label{eq:K2_dot}
\end{equation}
Next, by using the Jacobi property of the guiding-center Poisson bracket, we introduce the identity
\begin{eqnarray}
\left\{ \wt{\phi}_{{\rm gc}},\; \{ \wt{\Phi}_{{\rm gc}},\; \ov{G} \}_{{\rm gc}} \right\}_{{\rm gc}} & \equiv & \frac{1}{2}\; \left\{  \ov{G},\;
\{ \wt{\Phi}_{{\rm gc}},\; \wt{\phi}_{{\rm gc}}\}_{{\rm gc}} \right\}_{{\rm gc}} \nonumber \\
 &  &+\; \pd{}{\theta} \left( \frac{1}{2}\; \left\{ \wt{\Phi}_{{\rm gc}},\; \{ \wt{\Phi}_{{\rm gc}},\; \ov{G} \}_{{\rm gc}} \right\}_{{\rm gc}} \right),
\label{eq:id_PB}
\end{eqnarray}
where $\ov{G}$ is a gyroangle-independent function and $\wt{\phi}_{{\rm gc}} \equiv \partial\wt{\Phi}_{{\rm gc}}/\partial\theta$. Substituting 
(\ref{eq:K1_gc}) into (\ref{eq:K2_dot}) and using the identity (\ref{eq:id_PB}), the second-order gyrocenter kinetic equation (\ref{eq:K2_dot}) becomes
\[ (\dot{K}_{{\rm gy}})_{2} \;=\; -\; \left\{ K_{\rm gc},\; \frac{q^{2}}{2\Omega}\;\left\langle \{ \wt{\Phi}_{{\rm gc}},\; \wt{\phi}_{{\rm gc}}
\}_{{\rm gc}} \right\rangle \right\}_{{\rm gc}}. \]
The gyroangle-dependent part of the second-order kinetic energy, on the other hand, is expressed as
\begin{eqnarray}
\wt{K}_{2{\rm gy}} & \equiv & -\;\int \left[\; \frac{1}{\Omega}\,\frac{d_{{\rm gy}}\wt{K}_{1{\rm gy}}}{dt} \;+\; \frac{q}{\Omega} \left( \left\{ 
\wt{K}_{1{\rm gy}},\; \wt{\phi}_{{\rm gc}} \right\}_{{\rm gc}} \;-\; \left\langle \left\{ \wt{K}_{1{\rm gy}},\; \wt{\phi}_{{\rm gc}} \right\}_{{\rm gc}} \right\rangle \right) \;\right] d\theta \nonumber \\
 & = & -\;\frac{q}{\Omega}\; \frac{d_{{\rm gy}}\wt{\Phi}_{{\rm gc}}}{dt} \;+\; \cdots,
\label{eq:K2_sol}
\end{eqnarray}
where we have ignored terms of order $\Omega^{-2}$ and the gyroangle-independent part $\ov{K}_{2{\rm gy}}$ must be computed at
third order.

The gyrocenter kinetic energy $K_{{\rm gy}}$ is therefore expressed as
\begin{eqnarray}
K_{{\rm gy}} & = & K_{\rm gc} \;+\; q\;\wt{\phi}_{{\rm gc}} \;+\;
\frac{q}{\Omega}\,v_{\|}\;\bhat\bdot\nabla\wt{\Phi}_{{\rm gc}} \;-\;
\frac{q}{\Omega}\;  
\frac{d_{{\rm gy}}\wt{\Phi}_{{\rm gc}}}{dt} \;+\; \cdots \label{eq:Kgy_ivd} \\
 & = & K_{\rm gc} \;+\; q\;\wt{\phi}_{{\rm gc}}\;\left( 1 \;-\; \frac{q}{B}\;
   \pd{\langle\phi_{{\rm gc}}\rangle}{\mu} \right) \;-\; \frac{q}{\Omega}\; 
   \left( \pd{}{t} + \frac{c\bhat}{B}\btimes\nabla\langle\phi_{{\rm gc}}
   \rangle\bdot\nabla \right) \wt{\Phi}_{{\rm gc}} \;+\;\cdots, 
\nonumber
\end{eqnarray}
where terms of second order in $\Omega^{-1}$ have been omitted and the terms
of first order in $\Omega^{-1}$ associated with the gyroangle-independent part $\ov{K}_{2{\rm gy}}$ have not been
computed. While Parra \& Catto \cite{P_Catto} captured the first-order term correctly, their second-order
term includes $-\,(q/\Omega)\,\partial_{t}\wt{\Phi}_{{\rm gc}}$ only and ignores the second-order correction 
terms due to $\langle\phi_{{\rm gc}}\rangle$. In fact, 
Parra \& Catto \cite{P_Catto} systematically
ignore $\phi^{2}$-terms in their derivations of gyrokinetic variables except
in their revised quasineutrality condition. 

Lastly, the gyrocenter equation for $\dot{K}_{{\rm gy}}$ is expressed as
\begin{equation}
\dot{K}_{{\rm gy}} \;=\; -\;v_{\|}\;\bhat\bdot\nabla\left( q\,\langle\phi_{{\rm gc}}\rangle \;-\; \frac{q^{2}}{2\Omega}\;\left\langle \{ 
\wt{\Phi}_{{\rm gc}},\; \wt{\phi}_{{\rm gc}}\}_{{\rm gc}} \right\rangle \right),
\label{eq:Kgy_dot}
\end{equation}
which includes a nonlinear (quadratic) contribution to the parallel electric field generated by $\langle\phi_{{\rm gc}}\rangle$.

\section{\label{sec:gyro_Lie}Gyrocenter Lie-transform Approach}

The transformation from the extended guiding-center coordinates $Z_{\rm gc}^{\alpha} \equiv Z^{\alpha} = ({\bf X}, p_{\|}, \mu, \theta; w, t)$ to the gyrocenter coordinates $Z_{\rm gy}^{\alpha} \equiv \ov{Z}^{\alpha} = (\ov{{\bf X}}, \ov{p}_{\|}, \ov{\mu}, \ov{\theta}; \ov{w}, t)$ is expressed as an asymptotic expansion 
\begin{equation}
\ov{Z}^{\alpha} \;=\; Z^{\alpha} \;+\; \epsilon\,G_{1}^{\alpha} \;+\; \epsilon^{2} \left( G_{2}^{\alpha} \;+\; \frac{1}{2}\,G_{1}^{\beta}
\pd{G_{1}^{\alpha}}{Z^{\alpha}} \right) \;+\; \cdots,
\label{eq:Zgy_LT}
\end{equation}
where the $n$th-order generating vector field ${\sf G}_{n}$ is evaluated at order $\epsilon^{n}$ to eliminate gyroangle-dependence in the Hamiltonian. Here, the energy coordinate $w$ is introduced as the canonically-conjugate coordinate to time $t$. In Hamiltonian Lie-transform perturbation analysis (appropriate for electrostatic perturbations), the generating vector fields are expressed in terms of the extended guiding-center Poisson bracket 
$\{\;,\; \}_{{\rm gc}}$, which now includes the canonical pair $(w,t)$, as
\begin{equation}
G_{k}^{\alpha} \;\equiv\; \left\{ S_{k},\; Z^{\alpha} \right\}_{{\rm gc}},
\label{eq:Gk_Sk}
\end{equation}
where the functions $(S_{1}, S_{2}, \cdots)$ are known as the gyrocenter gauge functions (which are assumed to be explicitly gyroangle-dependent). We note that the time coordinate $t$ is unaffected by the time-dependent gyrocenter transformation if $G_{k}^{t} = \partial S_{k}/\partial w \equiv 0$ at all orders.

The extended guiding-center Hamiltonian is
\begin{equation}
H_{{\rm gc}} \;=\; \left( \frac{p_{\|}^{2}}{2m} \;+\; \mu\,B \;+\; q\,\langle\phi_{{\rm gc}}\rangle \;-\; w \right) \;+\; \epsilon\;
\left( q\;\wt{\phi}_{{\rm gc}} \right) \;\equiv\; H_{0{\rm gc}} \;+\; \epsilon\;H_{1{\rm gc}},
\label{eq:H_gc}
\end{equation}
where we have explicitly separated the gyroangle-dependent part $\wt{\phi}_{\rm gc}$ of the
guiding-center scalar potential $\phi_{\rm gc}$ as the perturbation that destroys the
guiding-center magnetic moment (i.e., $\{ \mu,\; H_{0{\rm gc}}\}_{{\rm gc}}
\equiv 0$ and $\{ \mu,\; H_{1{\rm gc}}\}_{{\rm gc}} \neq 0$). While the
separation adopted in (\ref{eq:H_gc}) is nonstandard \cite{Brizard_Hahm}, it is consistent with
the gcrV method presented in \S\,\ref{sec:gcrv}. Note that this separation appears when the electrostatic 
potential has a large-scale component $\gav{\phi_{\rm gc}}$ and a small-scale component $\wt{\phi}_{\rm gc}$, which satisfy the ordering 
\begin{equation}
q\,\wt{\phi}_{\rm gc} \;\ll\; T \;\sim\; q\,\langle\phi_{\rm gc}\rangle,
\label{eq:phi_ordering}
\end{equation}
where $T$ denotes the plasma temperature. This ordering is consistent with the generalized gyrokinetic ordering \cite{Dimits} $(\rho/\lambda_{\bot})\,
q\phi \ll T$, where the perpendicular gradient length scale $\lambda_{\bot}$ scales as $\lambda_{\bot} \sim \rho$ for the small-scale component 
$\wt{\phi}_{\rm gc}$ while it scales as $\lambda_{\bot} \gg \rho$ for the large-scale component $\gav{\phi_{\rm gc}}$.

The gyrocenter transformation (\ref{eq:Zgy_LT}) is chosen at each order so that the gyrocenter Hamiltonian
\begin{equation}
H_{{\rm gy}} \;=\; H_{0{\rm gy}} \;+\; \epsilon\,H_{1{\rm gy}} \;+\; \epsilon^{2}\,H_{2{\rm gy}} \;+\; \cdots
\label{eq:Hgy_def}
\end{equation}
is gyroangle-independent, where $H_{0{\rm gy}} \equiv H_{0{\rm gc}}$. According to Hamiltonian Lie-transform perturbation theory, the first-order and second-order gyrocenter Hamiltonians are
\begin{eqnarray}
H_{1{\rm gy}} & \equiv & H_{1{\rm gc}} \;-\; \left\{ S_{1},\; H_{0{\rm gc}} \right\}_{{\rm gc}} \;=\; H_{1{\rm gc}} \;-\; \left( \frac{d_{{\rm gy}}S_{1}}{dt} \;+\; \Omega\;\pd{S_{1}}{\theta} \right), \label{eq:Hgy_1def} \\
H_{2{\rm gy}} & = & -\; \{ S_{1},\; H_{1{\rm gc}} \}_{{\rm gc}} \;+\; \frac{1}{2}\; \left\{\, S_{1},\; \{ S_{1},\; H_{0{\rm gc}}\frac{}{}
\}_{{\rm gc}} \right\}_{{\rm gc}} \nonumber \\
 &  &-\; \left( \frac{d_{{\rm gy}}S_{2}}{dt} \;+\; \Omega\;\pd{S_{2}}{\theta} \right), \label{eq:Hgy_2def}
\end{eqnarray}
where we used $H_{2{\rm gc}} \equiv 0$. It is straightforward to extend the Lie transform approach to a nonuniform magnetic field, since the operator 
$d_{\rm gy}/dt = \{\;\cdot \; , \; H_{0 \rm gc}\}_{\rm gc}$ is valid in general magnetic geometry, with the guiding-center Poisson bracket suitably generalized for nonuniform magnetic fields [see (\ref{eq:gcPB_general})].

\subsection{First-order analysis}

At first order, since $H_{1{\rm gc}} = q\,\wt{\phi}_{\rm gc}$ and $\langle H_{1{\rm gc}}\rangle = 0$, the expression for the gyrocenter Hamiltonian is simply
\begin{equation}
H_{1{\rm gy}} \;\equiv\; 0,
\label{eq:Hgy_1}
\end{equation}
while the first-order gauge function $S_{1}$ is the solution of the gyroangle-dependent equation
\begin{equation}
\frac{d_{{\rm gy}}S_{1}}{dt} \;+\; \Omega\;\pd{S_{1}}{\theta} \;=\; q\,\wt{\phi}_{{\rm gc}}.
\label{eq:dS1dt_def}
\end{equation}
The reader should not be alarmed by (\ref{eq:Hgy_1}) and remember that the gyroangle-averaged scalar potential $\langle\phi_{{\rm gc}}\rangle$ appears in the gyroangle-independent perturbed guiding-center Hamiltonian $H_{0{\rm gc}}$ in (\ref{eq:H_gc}). 

Up to second order in $\Omega^{-1}$, the solution for $S_{1}$ is
\begin{equation}
S_{1} \;=\; \frac{q}{\Omega}\;\wt{\Phi}_{{\rm gc}} \;-\; \frac{q}{\Omega^{2}}\;\frac{d_{{\rm gy}}\wt{\Phi}^{(2)}_{{\rm gc}}}{dt} \;+\; \cdots
\label{eq:S1_gy}
\end{equation}
where $\wt{\Phi}^{(k+1)}_{{\rm gc}} \equiv \int\, \wt{\Phi}^{(k)}_{{\rm gc}}\,d\theta$, with $\wt{\Phi}^{(1)}_{{\rm gc}} \equiv \wt{\Phi}_{{\rm gc}}$ and
$\wt{\Phi}^{(0)}_{{\rm gc}} \equiv \wt{\phi}_{{\rm gc}}$.

\subsection{Second-order analysis}

At second order, the expression for the gyrocenter Hamiltonian is
\begin{equation}
H_{2{\rm gy}} \;\equiv\; -\;\frac{q^{2}}{2\Omega}\;\left\langle \left\{ \left( \wt{\Phi}_{{\rm gc}} \;-\; \frac{1}{\Omega}\,
\frac{d_{\rm gy}\wt{\Phi}_{\rm gc}^{(2)}}{dt} \right),\; \wt{\phi}_{{\rm gc}} \right\}_{{\rm gc}}\right\rangle,
\label{eq:Hgy_2}
\end{equation}
while the second-order gauge function $S_{2}$ is the solution of the gyroangle-dependent equation
\begin{equation}
\frac{d_{{\rm gy}}S_{2}}{dt} \;+\; \Omega\;\pd{S_{2}}{\theta} \;=\; -\; \frac{q^{2}}{2\Omega}\; \left( \left\{ \wt{\Phi}_{{\rm gc}},\; 
\wt{\phi}_{{\rm gc}} \right\}_{{\rm gc}} \;-\; \left\langle \left\{ \wt{\Phi}_{{\rm gc}},\; \wt{\phi}_{{\rm gc}} \right\}_{{\rm gc}} 
\right\rangle \right),
\label{eq:S2_def}
\end{equation}
where terms of order $\Omega^{-2}$ were omitted on the right side of (\ref{eq:S2_def}). Up to second order in $\Omega^{-1}$, the solution for $S_{2}$ is
\begin{equation}
S_{2} \;=\; -\; \frac{q^{2}}{2\Omega^{2}} \int \left( \left\{ \wt{\Phi}_{{\rm gc}},\; \wt{\phi}_{{\rm gc}} \right\}_{{\rm gc}} 
\;-\; \left\langle \left\{ \wt{\Phi}_{{\rm gc}},\; \wt{\phi}_{{\rm gc}} \right\}_{{\rm gc}} \right\rangle \right) d\theta.
\label{eq:S2_gy}
\end{equation}

\subsection{Gyrocenter Coordinates}

One of the main advantages of the gyrocenter Lie-transform approach is that the gyrocenter phase-space transformation is expressed solely in terms of the scalar fields $(S_{1}, S_{2}, \cdots)$. The gyrocenter phase-space coordinates are constructed from the asymptotic expansion
\begin{eqnarray}
\ov{Z}^{\alpha} & = & Z^{\alpha} \;+\; \epsilon\,\left\{ S_{1},\; Z^{\alpha}\right\}_{{\rm gc}} \;+\; \epsilon^{2} \left( \left\{ S_{2},\; 
Z^{\alpha}\right\}_{{\rm gc}} \;+\; \frac{1}{2}\,\left\{ S_{1},\; \{S_{1},\; Z^{\alpha}\}_{{\rm gc}} \right\}_{{\rm gc}} \right) \;+\; \cdots, 
\label{eq:Zgy_S} \\
 & = & Z^{\alpha} \;+\; \epsilon\,\frac{q}{\Omega}\;\left\{ \left( \wt{\Phi}_{{\rm gc}} \;-\; \frac{1}{\Omega}\;\frac{d_{{\rm gy}}
\wt{\Phi}^{(2)}_{{\rm gc}}}{dt}\right),\; Z^{\alpha}\right\}_{{\rm gc}} \nonumber \\
 &  &+\; \epsilon^{2} \left( \left\{ S_{2},\; Z^{\alpha}\right\}_{{\rm gc}} \;+\; \frac{q^{2}}{2\Omega^{2}}\,\left\{ \wt{\Phi}_{{\rm gc}},\; \{\wt{\Phi}_{{\rm gc}},\; Z^{\alpha}\}_{{\rm gc}} \right\}_{{\rm gc}} \right) \;+\; \cdots, \nonumber 
\end{eqnarray}
where we substituted the expression (\ref{eq:S1_gy}) for $S_{1}$ while the expression (\ref{eq:S2_gy}) for $S_{2}$ will be used only when needed.
Hence, the gyrocenter position $\ov{{\bf X}}$ is
\begin{equation}
\ov{{\bf X}} \;=\; {\bf X} \;+\; \epsilon\,\frac{q}{\Omega}\;\left\{ \wt{\Phi}_{{\rm gc}},\; {\bf X} \right\}_{{\rm gc}} \;+\; \cdots \;=\; {\bf X} 
\;-\; \epsilon\,\frac{c\bhat}{B\Omega}\btimes\nabla\wt{\Phi}_{{\rm gc}} \;+\; \cdots,
\label{eq:Rgy_S}
\end{equation}
the gyrocenter parallel momentum $\ov{p}_{\|}$ is
\begin{equation}
\ov{p}_{\|} \;=\; p_{\|} \;+\; \epsilon\,\frac{q}{\Omega}\;\left\{ \wt{\Phi}_{{\rm gc}},\; p_{\|} \right\}_{{\rm gc}} \;+\; \cdots \;=\; p_{\|} \;+\; 
\epsilon\,\frac{q\bhat}{\Omega}\bdot\nabla\wt{\Phi}_{{\rm gc}} \;+\; \cdots,
\label{eq:vgy_S}
\end{equation}
the gyrocenter gyroangle $\ov{\theta}$ is
\begin{equation}
\ov{\theta} \;=\; \theta \;+\; \epsilon\,\frac{q}{\Omega}\;\left\{ \wt{\Phi}_{{\rm gc}},\; \theta \right\}_{{\rm gc}} \;+\; \cdots \;=\; \theta \;-\; 
\epsilon\,\frac{q}{B}\;\pd{\wt{\Phi}_{{\rm gc}}}{\mu} \;+\; \cdots,
\label{eq:thetagy_S}
\end{equation}
the gyrocenter magnetic moment $\ov{\mu}$ is
\begin{eqnarray}
\ov{\mu} & = & \mu \;+\; \epsilon\,\frac{q}{\Omega}\;\left\{ \left( \wt{\Phi}_{{\rm gc}} \;-\; \frac{1}{\Omega}\;\frac{d_{{\rm gy}}
\wt{\Phi}^{(2)}_{{\rm gc}}}{dt}\right),\; \mu \right\}_{{\rm gc}} \nonumber \\
 &  &+\; \epsilon^{2}\,\left( \left\{ S_{2},\; \mu \right\}_{{\rm gc}} \;+\; 
\frac{q^{2}}{2\Omega^{2}}\,\left\{ \wt{\Phi}_{{\rm gc}},\; \{\wt{\Phi}_{{\rm gc}},\; \mu \}_{{\rm gc}} \right\}_{{\rm gc}} \right) \;+\; \cdots
\nonumber \\
 & = & \mu \;+\; \epsilon\,\frac{q}{B} \left( \wt{\phi}_{{\rm gc}} \;-\; \frac{1}{\Omega}\;\frac{d_{{\rm gy}}\wt{\Phi}_{{\rm gc}}}{dt} \right) \;+\; 
\epsilon^{2}\,\frac{q^{2}}{2B\Omega}\; \left\langle \left\{ \wt{\Phi}_{{\rm gc}},\; \wt{\phi}_{{\rm gc}} \right\}_{{\rm gc}} \right\rangle \;+\; \cdots,
\label{eq:mugy_S}
\end{eqnarray}
where the expression (\ref{eq:S2_gy}) for $S_{2}$ was used in obtaining the last expression, and the gyrocenter energy coordinate $\ov{w}$ is
\begin{equation}
\ov{w} \;=\; w \;-\; \epsilon\;\frac{q}{\Omega}\,\pd{\wt{\Phi}_{\rm gc}}{t} \;+\; \cdots. 
\label{eq:wgy_S}
\end{equation}

The gyrocenter kinetic energy $K_{{\rm gy}}$ is
\begin{eqnarray}
K_{{\rm gy}} & = & K_{\rm gc} \;+\; \epsilon\,\frac{q}{\Omega}\;\left\{ \left( \wt{\Phi}_{{\rm gc}} \;-\; \frac{1}{\Omega}\;\frac{d_{{\rm gy}}
\wt{\Phi}^{(2)}_{{\rm gc}}}{dt}\right),\; K_{\rm gc} \right\}_{{\rm gc}} \nonumber \\
 &  &+\; \epsilon^{2}\,\left( \left\{ S_{2},\; K_{\rm gc} \right\}_{{\rm gc}} \;+\; \frac{q^{2}}{2\Omega^{2}}\,\left\{ \wt{\Phi}_{{\rm gc}},\; \{\wt{\Phi}_{{\rm gc}},\; K_{\rm gc} \}_{{\rm gc}} \right\}_{{\rm gc}} \right) \;+\; \cdots \nonumber \\
 & = & K_{\rm gc} \;+\; \epsilon\,q\,\wt{\phi}_{{\rm gc}}\;\left( 1 \;-\; \frac{q}{B}\;\pd{\langle\phi_{{\rm gc}}\rangle}{\mu} \right) \;-\; 
\epsilon\,\frac{q}{\Omega}\;\left( \pd{}{t} \;+\; \frac{c\bhat}{B}\btimes\nabla\langle\phi_{{\rm gc}}\rangle
\bdot\nabla \right) \wt{\Phi}_{{\rm gc}} \nonumber \\
 &  &+\; \epsilon^{2}\,\frac{q^{2}}{2\Omega}\,\left\langle \left\{ \wt{\Phi}_{{\rm gc}},\; \wt{\phi}_{{\rm gc}}
\right\}_{{\rm gc}} \right\rangle \;+\; \cdots,
\label{eq:Kgy_S}
\end{eqnarray}
where terms of second order in $\Omega^{-1}$ have been omitted. One can see that the second-order gyro-independent part of the kinetic energy
$\ov{K}_{2 \rm gy}$ appears naturally in the Lie-transform approach, whereas it would require an extensive computation to obtain it in the guiding-center recursive Vlasov approach (we have skipped this computation, see \S\,\ref{subsec:kin_energy_it} for details). Note that the gyrocenter kinetic energy (\ref{eq:Kgy_S}) can be exactly expressed as
\begin{equation}
K_{{\rm gy}} \;\equiv\; \frac{\ov{p}_{\|}^{2}}{2m} \;+\; \ov{\mu}\,B,
\label{eq:Kgy_gy}
\end{equation}
when the definitions (\ref{eq:vgy_S}) and (\ref{eq:mugy_S}) for $\ov{p}_{\|}$ and $\ov{\mu}$ are used.

Lastly, the Jacobian for the gyrocenter transformation is
\begin{equation}
{\cal J}_{{\rm gy}} \;=\; {\cal J}_{\rm gc} \;-\; \epsilon\,\pd{}{Z^{\alpha}} \left( {\cal J}_{\rm gc}\;\left\{ S_{1},\; 
Z^{\alpha} \right\}_{{\rm gc}} \right) \;+\; \cdots \;\equiv\; {\cal J}_{\rm gc}, 
\label{eq:Jac_gy}
\end{equation}
where ${\cal J}_{\rm gc}$ is a constant in a uniform magnetic field. This result comes from the fact that
\[ \pd{}{Z^{\alpha}} \left( {\cal J}_{\rm gc}\;\left\{ S_{1},\; Z^{\alpha} \right\}_{{\rm gc}} \right) \;=\; -\;\pd{}{Z^{\alpha}} \left( 
{\cal J}_{\rm gc}\,J^{\alpha\beta}\; \pd{S_{1}}{Z^{\beta}} \right) \;=\; -\;{\cal J}_{\rm gc}\,J^{\alpha\beta}\;
\frac{\partial^{2}S_{1}}{\partial Z^{\alpha}\partial Z^{\beta}} \;\equiv\; 0, \]
which follows from the antisymmetry of the guiding-center Poisson tensor $J^{\alpha\beta}$ and the Liouville identities $\partial_{\alpha}({\cal J}_{\rm gc}\,J^{\alpha\beta}) = 0$. Note that (\ref{eq:Jac_gy}) is true to all orders in $\epsilon$.

\section{\label{sec:gyro_Vlasov_eq}Gyrokinetic Vlasov Equation}

We now derive the gyrokinetic Vlasov equation (\ref{eq:Vlasov_eq}) by the gcrV method and show how the gyrocenter phase-space transformation (\ref{eq:Zgy_S}) generated by the Lie-transform scalar fields $(S_{1}, S_{2}, \cdots)$ are involved in the transformation from the particle Vlasov distribution $f$ and the gyrocenter Vlasov distribution $\ov{F}$.

\subsection{Recursive Vlasov Derivation}

First, the particle Vlasov distribution $f$ is expressed in terms of the guiding-center Vlasov distribution $F$ by the push-forward operation $F \equiv
{\sf T}_{{\rm gc}}^{-1}f = F_{0} + \epsilon\,F_{1} + \epsilon^{2}\,F_{2} + \cdots$, where $F_{k} \equiv {\sf T}_{{\rm gc}}^{-1}f_{k}$. At zeroth, first, and second orders, we therefore have
\begin{equation}
\left. \begin{array}{rcl}
0 & = & L_{0{\rm gc}}F_{0} \\
0 & = & L_{1{\rm gc}}F_{0} \;+\; L_{0{\rm gc}}F_{1} \\
0 & = & L_{1{\rm gc}}F_{1} \;+\; L_{0{\rm gc}}F_{2}
\end{array} \right\},
\label{eq:gyVE_012}
\end{equation}
where the guiding-center operators $L_{0{\rm gc}}$ and $L_{1{\rm gc}}$ are defined in (\ref{eq:L0_gc}) and (\ref{eq:L1_gc}). 

At the zeroth order, the Vlasov equation $L_{0{\rm gc}}F_{0} = 0$ implies that $F_{0}$ is independent of the gyroangle $\theta$. At first order, the gyroangle-independent part of $L_{1{\rm gc}}F_{0}$ yields
\begin{equation}
0 \;=\; \frac{d_{{\rm gy}}F_{0}}{dt} \;=\; \frac{d_{{\rm gc}}F_{0}}{dt} \;+\; q\;\left\{ F_{0},\; \langle \phi_{{\rm gc}}\rangle \right\}_{{\rm gc}},
\label{eq:gyVE_1}
\end{equation}
while the gyroangle-dependent part yields a solution for $\wt{F}_{1}$:
\begin{equation}
\wt{F}_{1} \;\equiv\; -\;L_{0{\rm gc}}^{-1}\left( L_{1{\rm gc}}\,F_{0} \frac{}{}\right) \;=\; -\;\frac{q}{\Omega}\; \left\{ F_{0},\; \wt{\Phi}_{{\rm gc}} \right\}_{{\rm gc}}.
\label{eq:gyF_1}
\end{equation}
The solution for the gyroangle-independent part $\ov{F}_{1} \equiv F_{1} - \wt{F}_{1}$ must come from the second-order analysis.

At second order, the gyroangle-independent part of $L_{1{\rm gc}}F_{1}$ yields
\begin{eqnarray}
0  & = & \frac{d_{{\rm gy}}\ov{F}_{1}}{dt} \;+\; q\;\left\langle\left\{ \wt{F}_{1},\; \wt{\phi}_{{\rm gc}} \right\}_{{\rm gc}} \right\rangle \nonumber \\ 
   & = & \frac{d_{{\rm gy}}\ov{F}_{1}}{dt} \;-\; \left\{ F_{0},\; \frac{q^{2}}{2\Omega}\;\left\langle \left\{ \wt{\Phi}_{{\rm gc}},\; \wt{\phi}_{{\rm gc}} \right\}_{{\rm gc}} \right\rangle \right\}_{{\rm gc}},
\label{eq:gyVE_2}
\end{eqnarray}
while the gyroangle-dependent part yields a solution for $\wt{F}_{2}$:
\begin{eqnarray}
\wt{F}_{2} & = & -\; \frac{1}{\Omega}\; \int \left[\; \frac{d_{{\rm gy}}\wt{F}_{1}}{dt} \;+\; q\;\left\{ \ov{F}_{1},\; \wt{\phi}_{{\rm gc}} 
\right\}_{{\rm gc}} \right. \nonumber \\
 &  &\left.\hspace*{0.6in}+\; q \left( \left\{ \wt{F}_{1},\; \wt{\phi}_{{\rm gc}} \right\}_{{\rm gc}} \;-\; \left\langle\left\{ \wt{F}_{1},\; 
\wt{\phi}_{{\rm gc}} \right\}_{{\rm gc}} \right\rangle \right) \;\right] d\theta \nonumber \\
 & = & \frac{q}{\Omega^{2}}\; \left\{ F_{0},\; \frac{d_{{\rm gy}}\wt{\Phi}_{{\rm gc}}^{(2)}}{dt} \right\}_{{\rm gc}} \;-\;
\frac{q}{\Omega}\; \left\{ \ov{F}_{1},\; \wt{\Phi}_{{\rm gc}} \right\}_{{\rm gc}} \;+\; \frac{q^{2}}{2\,\Omega^{2}}\; \left\{ \left\{ F_{0},\;
\wt{\Phi}_{{\rm gc}} \right\}_{{\rm gc}},\; \wt{\Phi}_{{\rm gc}} \right\}_{{\rm gc}} \nonumber \\
 &  &+\; \left\{ F_{0},\; \left[\; \frac{q^{2}}{2\,\Omega^{2}}\; \int \left( \left\{ \wt{\Phi}_{{\rm gc}},\; \wt{\phi}_{{\rm gc}} \right\}_{{\rm gc}}
\;-\; \left\langle \left\{ \wt{\Phi}_{{\rm gc}},\; \wt{\phi}_{{\rm gc}} \right\}_{{\rm gc}} \right\rangle \right) d\theta \;\right] \right\}_{{\rm gc}}.
\label{eq:gyF_2}
\end{eqnarray}

\subsection{Gyrocenter Pull-back Operator}

We can combine the recursive solutions (\ref{eq:gyF_1}) and (\ref{eq:gyF_2}) for the guiding-center Vlasov distribution to obtain the following expansion
\begin{eqnarray}
F & \equiv & F_{0} \;+\; \epsilon\,F_{1} \;+\; \epsilon^{2}\,F_{2} \;+\; \cdots \;=\; \left( F_{0} \;+\; \epsilon\,\ov{F}_{1} \;+\; \cdots \right) \;+\;
\left( \epsilon\,\wt{F}_{1} \;+\; \epsilon^{2}\,\wt{F}_{2} \;+\; \cdots \right) \nonumber \\
 & = & \ov{F} \;+\; \epsilon\; \frac{q}{\Omega} \left\{ \left( \wt{\Phi}_{{\rm gc}} \;-\; \frac{1}{\Omega}\,
\frac{d_{{\rm gy}}\wt{\Phi}_{{\rm gc}}^{(2)}}{dt} \right),\; F_{0} \right\}_{{\rm gc}} \nonumber \\
 &  &+\; \epsilon^{2}\; \frac{q}{\Omega} \left\{ 
\wt{\Phi}_{{\rm gc}},\; \ov{F}_{1} \right\}_{{\rm gc}} \;+\; \frac{q^{2}\epsilon^{2}}{2\,\Omega^{2}}\; \left\{ \wt{\Phi}_{{\rm gc}},\; \left\{ \wt{\Phi}_{{\rm gc}},\; F_{0} \right\}_{{\rm gc}} \right\}_{{\rm gc}} \nonumber \\
 &  &+\; \epsilon^{2}\; \left\{ F_{0},\; \left[\; \frac{q^{2}}{2\,\Omega^{2}}\; \int \left( \left\{ \wt{\Phi}_{{\rm gc}},\; \wt{\phi}_{{\rm gc}} 
\right\}_{{\rm gc}} \;-\; \left\langle \left\{ \wt{\Phi}_{{\rm gc}},\; \wt{\phi}_{{\rm gc}} \right\}_{{\rm gc}} \right\rangle \right) d\theta \;\right] \right\}_{{\rm gc}}, 
\label{eq:F_Fbar}
\end{eqnarray}
where $\ov{F} \equiv F_{0} + \epsilon\,\ov{F}_{1} + \epsilon^{2}\,\ov{F}_{2} + \cdots$ defines the {\it gyrocenter} Vlasov distribution and we used the identity (\ref{eq:id_PB}) for the last two terms. Using the gyrocenter scalar fields $(S_{1}, S_{2}, \cdots)$, the relation (\ref{eq:F_Fbar}) between the guiding-center Vlasov distribution $F$ and the gyrocenter Vlasov distribution $\ov{F}$ can also be expressed as
\begin{eqnarray}
F & = & \ov{F} \;+\; \epsilon \left\{ S_{1},\; \ov{F} \right\}_{{\rm gc}} \;+\; \frac{\epsilon^{2}}{2} \left\{ S_{1},\; \{ S_{1},\; \ov{F} \} \frac{}{}\right\}_{{\rm gc}} \;+\; \epsilon^{2} \left\{ S_{2},\; \ov{F} \right\}_{{\rm gc}} \;+\; \cdots  \nonumber \\
 & \equiv & {\sf T}_{{\rm gy}}\,\ov{F},
\label{eq:FFbar_S}
\end{eqnarray}
where the guiding-center Vlasov distribution $F \equiv {\sf T}_{{\rm gy}}\,\ov{F}$ is represented as the gyrocenter pull-back of the gyrocenter Vlasov distribution $\ov{F}$. We immediately see that the gcrV method has generated a solution that is exactly expressed in terms of the gyrocenter pull-back operator ${\sf T}_{{\rm gy}}$. The physical interpretation of the gyrocenter pull-back operation is therefore given in terms of the time integration of the guiding-center Vlasov equation over the fast gyromotion time scale.

The gyrocenter pull-back also generates the standard decomposition of the perturbed particle Vlasov distribution in terms of its adiabatic and non-adiabatic parts as follows. Up to first order in $\epsilon$, we find
\begin{eqnarray} 
f_{1} & = & {\sf T}_{{\rm gc}} \left( \ov{F}_{1} \;+\; \frac{q}{\Omega} \left\{ \wt{\Phi}_{{\rm gc}},\; F_{0} \right\}_{{\rm gc}} \right) \nonumber \\
 & = & {\sf T}_{{\rm gc}} \left[\; \ov{F}_{1} \;+\; q\;\left( \pd{F_{0}}{E} \;+\; \frac{1}{B}\,\pd{F_{0}}{\mu} \right) 
\wt{\phi}_{{\rm gc}} \;+\; \cdots \;\right],
\label{eq:f1_F1}
\end{eqnarray}
where higher-order corrections have been neglected and the guiding-center Poisson bracket is now expressed in terms of the guiding-center coordinates 
$({\bf X},E,\mu,\theta)$ with the guiding-center energy $E$ replacing the guiding-center parallel kinetic momentum $p_{\|}$. Next, we introduce the decomposition
\begin{equation}
\ov{F}_{1} \;=\; q\,\langle\phi_{{\rm gc}}\rangle\;\pd{F_{0}}{E} \;+\; \ov{G}_{1},
\label{eq:FG_1}
\end{equation}
where the first term represents the {\it adiabatic} contribution to $\ov{F}_{1}$ and $\ov{G}_{1}$ represents its {\it non-adiabatic} contribution. Lastly, we use the identity (\ref{eq:Tphi_gc}) to obtain the relation \cite{Rutherford}
\begin{equation}
f_{1} \;=\; q\,\phi\;\left( \pd{f_{0}}{E} \;+\; \frac{1}{B}\,\pd{f_{0}}{\mu} \right) \;+\; {\sf T}_{{\rm gc}} \left( \ov{G}_{1} \;-\; \frac{q}{B}\,
\langle\phi_{{\rm gc}}\rangle\;\pd{F_{0}}{\mu} \right),
\label{eq:f1}
\end{equation}
where $f_{0} \equiv {\sf T}_{{\rm gc}}F_{0}$. Here, we note that the adiabatic contribution naturally separates into a particle part (involving $\phi$) and a guiding-center part (involving ${\sf T}_{{\rm gc}}\,\langle\phi_{{\rm gc}}\rangle$).

\subsection{Gyrokinetic Vlasov Equation}

By combining the gyrocenter Vlasov equations (\ref{eq:gyVE_1}) and (\ref{eq:gyVE_2}) we obtain the nonlinear gyrokinetic Vlasov equation
\begin{eqnarray}
0 & = & \frac{d_{{\rm gc}}\ov{F}}{dt} \;+\; \left\{ \ov{F},\; q\,\Psi_{{\rm gy}} \right\}_{{\rm gc}} \label{eq:gyVE} \\
 & = & \frac{d_{\rm gy}\ov{F}}{dt} \;-\; \left\{ \ov{F},\; \frac{q^{2}}{2\,\Omega} \left\langle \left\{ \left( \wt{\Phi}_{\rm gc} - \frac{1}{\Omega}\,\frac{d_{\rm gy}\wt{\Phi}^{(2)}_{\rm gc}}{dt} \right),\; \wt{\phi}_{{\rm gc}} \right\}_{{\rm gc}} \right\rangle \right\}_{\rm gc}, \nonumber
\end{eqnarray}
where the effective gyrocenter potential
\begin{eqnarray}
\Psi_{{\rm gy}} & \equiv & \langle\phi_{{\rm gc}}\rangle \;-\; \frac{q}{2\,\Omega}\;\left\langle 
\left\{ \left( \wt{\Phi}_{{\rm gc}} \;-\; \frac{1}{\Omega}\,\frac{d_{\rm gy}\wt{\Phi}_{\rm gc}^{(2)}}{dt} \right),\; 
\wt{\phi}_{{\rm gc}} \right\}_{{\rm gc}} \right\rangle \nonumber \\
 & = & \langle\phi_{{\rm gc}}\rangle \;-\; \frac{q}{2\,\Omega}\;\left\langle \left\{ \wt{\Phi}_{{\rm gc}},\; 
\wt{\phi}_{{\rm gc}} \right\}_{{\rm gc}} \right\rangle \;-\; \frac{q}{2\,\Omega^{2}}\;\left\langle \left\{ \,\frac{d_{\rm gy}\wt{\Phi}_{\rm gc}}{dt},\;
\wt{\Phi}_{{\rm gc}} \right\}_{{\rm gc}} \right\rangle
\label{eq:Psi_gy}
\end{eqnarray} 
contains nonlinear ponderomotive corrections to the linear scalar potential. 

We note that in standard gyrokinetic theory \cite{Dubin}, the polarization-drift correction (involving $d_{\rm gy}\wt{\Phi}_{\rm gc}/dt$) is omitted and only the second term is retained in (\ref{eq:Psi_gy}). The gyrokinetic Vlasov equation (\ref{eq:Psi_gy}) describes the time evolution of the gyroangle-independent gyrocenter Vlasov distribution 
$\ov{F}(\ov{{\bf X}}, \ov{p}_{\|}, t;\, \ov{\mu})$ in a $4+1$ reduced phase space, where the gyrocenter Hamilton's equations (in a uniform magnetic field) are 
\begin{equation}
\dot{\ov{{\bf X}}} \;\equiv\; \ov{v}_{\|}\,\bhat \;+\; \frac{c\bhat}{B}\btimes\ov{\nabla}\Psi_{\rm gy} \;\;\;{\rm and}\;\;\; 
\dot{\ov{p}}_{\|} \;\equiv\; -\,q\,\bhat\bdot\ov{\nabla}\Psi_{\rm gy},
\label{eq:gyro_Ham}
\end{equation}
and the gyrocenter magnetic moment $\ov{\mu}$ is an invariant.

A common approximation for the gyrokinetic Vlasov equation (\ref{eq:gyVE}) is obtained by writing it in truncated form as $d_{\rm gy}\ov{F}/dt \equiv 0$ and then expressing, first, the gyrocenter Vlasov distribution function as $\ov{F} = F_{0} + \ov{F}_{1}$ and, second, using the decomposition 
(\ref{eq:FG_1}) to obtain a gyrokinetic equation for the nonadiabatic part $\ov{G}_{1}$. The electrostatic version of the Frieman-Chen gyrokinetic equation \cite{Frieman_Chen} is thus obtained from the truncated equation 
\begin{eqnarray}
\frac{d_{{\rm gy}}\ov{F}_{1}}{dt} & = & -\;\frac{d_{{\rm gy}}F_{0}}{dt} \;=\; -\;q \left\{ F_{0},\; \langle\phi_{{\rm gc}}\rangle\right\}_{{\rm gc}} \nonumber \\
 & = & q\,\pd{F_{0}}{\ov{E}}\;\left( \frac{d_{{\rm gc}}}{dt} \;-\; \pd{}{t} \right)\langle\phi_{{\rm gc}}\rangle \;-\; \frac{c\bhat}{B}\btimes
\ov{\nabla}\langle\phi_{{\rm gc}}\rangle\bdot\ov{\nabla} F_{0}, 
\label{eq:FC_1}
\end{eqnarray}
where the gyrocenter kinetic-energy coordinate $\ov{E}$ [see (\ref{eq:Kgy_gy})] is used instead of $\ov{p}_{\|}$ and the background distribution $F_{0}$ satisfies the guiding-center Vlasov equation $d_{\rm gc}F_{0}/dt \equiv 0$. Next, we introduce the decomposition (\ref{eq:FG_1}), where we write
\[ \frac{d_{{\rm gy}}}{dt} \left( q\,\langle\phi_{{\rm gc}}\rangle\;\pd{F_{0}}{\ov{E}} \right) \;=\; q\;
\frac{d_{{\rm gc}}\langle\phi_{{\rm gc}}\rangle}{dt}\;\pd{F_{0}}{\ov{E}} \;+\; \cdots, \]
to obtain the electrostatic Frieman-Chen gyrokinetic equation 
\begin{equation}
\frac{d_{{\rm gy}}\ov{G}_{1}}{dt} \;=\; -\; \left( q\,\pd{\langle\phi_{{\rm gc}}\rangle}{t}\;\pd{}{\ov{E}} \;+\; \frac{c\bhat}{B}\btimes\ov{\nabla}
\langle\phi_{{\rm gc}}\rangle\bdot\ov{\nabla} \right) F_{0}.
\label{eq:FC_2}
\end{equation}
While this equation offers great simplicity for many practical applications, it also suffers from several deficiencies \cite{Brizard_Hahm} (e.g., it lacks energy conservation when combined with the gyrokinetic Poisson equation) which limit its use in numerical simulations of electrostatic plasma turbulence. 

\section{\label{sec:Poisson}Gyrokinetic Poisson Equation}

When the nonlinear gyrokinetic Vlasov equation (\ref{eq:gyVE}) is combined with the gyrokinetic version of the Poisson equation, we obtain a set of energy-conserving equations that can be used for numerical simulations of electrostatic plasma turbulence \cite{Brizard_Hahm}.

The gyrokinetic Poisson equation is expressed as a moment of the gyrocenter Vlasov distribution $\ov{F}$ through a sequence of phase-space transformations from particle to guiding-center to gyrocenter phase spaces:
\begin{equation}
-\;\frac{\nabla^{2}\phi({\bf r})}{4\pi} \;=\; q\;\int d^{6}z\; f\; \delta^{3} \;=\; q\;\int d^{6}Z\; F\; \delta_{\rm gc}^{3} \;\equiv\;
q\;\int d^{6}\ov{Z}\; \ov{F}\; \left\langle {\sf T}_{\rm gy}^{-1}\delta_{\rm gc}^{3}\right\rangle,
\label{eq:gyro_Poisson}
\end{equation}
where summation over particle species is implied, $\delta^{3} = \delta^{3}({\bf x} - {\bf r})$ implies that only particles located at the field position ${\bf x} = {\bf r}$ contribute to the scalar field $\phi({\bf r})$, and $\delta_{\rm gc}^{3} \equiv {\sf T}_{\rm gc}^{-1}\delta^{3} = \delta^{3}({\bf X} + \vb{\rho}_{\rm gc} - {\bf r})$ is expressed in terms of the guiding-center gyroradius vector $\vb{\rho}_{\rm gc}$. The last expression in (\ref{eq:gyro_Poisson}) involves the gyrocenter push-forward operation
\[ {\sf T}_{\rm gy}^{-1}g \;=\; g \;-\; \epsilon\;\{S_{1},\; g\}_{\rm gc} \;-\; \epsilon^{2} \left( \{S_{2},\; g\}_{\rm gc} \;-\; \frac{1}{2}\,
\left\{S_{1},\frac{}{} \{S_{1},\; g\}_{\rm gc}\right\}_{\rm gc} \right) \;+\; \cdots, \]
where the generating scalar fields $S_{1}$ and $S_{2}$ are defined in (\ref{eq:S1_gy}) and (\ref{eq:S2_gy}). When the push-forward operator is applied to $\delta_{\rm gc}^{3}$, we obtain 
\begin{eqnarray}
\left\langle{\sf T}_{\rm gy}^{-1}\delta_{\rm gc}^{3}\right\rangle & = & \langle \delta_{\rm gc}^{3}\rangle \;+\; \epsilon^{2}\;\frac{q^{2}}{2\Omega^{2}}\;\left\langle \left\{ \wt{\Phi}_{\rm gc},\frac{}{} \left\{ \wt{\Phi}_{\rm gc},\; \langle\delta_{\rm gc}^{3}\rangle \right\}_{\rm gc} \right\}_{\rm gc} \right\rangle
\;+\; \cdots \nonumber \\
 &  &-\; \epsilon\;\frac{q}{\Omega}\,\left\langle \left\{ \left(\wt{\Phi}_{\rm gc} - \frac{1}{\Omega}\,\frac{d_{\rm gy}\wt{\Phi}_{\rm gc}^{(2)}}{dt} \right),\; \wt{\delta}_{\rm gc}^{3} \right\}_{\rm gc} \right\rangle \;+\; \cdots, 
\label{eq:push_delta}
\end{eqnarray}
where $\wt{\delta}_{\rm gc}^{3} \equiv \delta_{\rm gc}^{3} - \langle \delta_{\rm gc}^{3}\rangle = \vb{\rho}_{\rm gc}\bdot\nabla\delta^{3} + \cdots$ denotes the gyroangle-dependent part of $\delta_{\rm gc}^{3}$. We note that polarization effects enter into the gyrokinetic Poisson equation 
(\ref{eq:gyro_Poisson}) through the term $\langle{\sf T}_{\rm gy}^{-1}\delta_{\rm gc}^{3}\rangle \equiv \langle{\sf T}_{\rm gy}^{-1}({\sf T}_{\rm gc}^{-1}\delta^{3})\rangle$: guiding-center polarization enters through the difference $\langle{\sf T}_{\rm gc}^{-1}\delta^{3}\rangle - \delta^{3}$ while gyrocenter polarization enters through the difference $\langle{\sf T}_{\rm gy}^{-1}\delta_{\rm gc}^{3}\rangle - \langle\delta_{\rm gc}^{3}\rangle$. We further note that (\ref{eq:push_delta}) may also be obtained from the functional derivative of the effective gyrocenter potential (\ref{eq:Psi_gy}):
\begin{equation}
\frac{\delta}{\delta\phi}\Psi_{\rm gy}[\phi] \;\equiv\; \left\langle {\sf T}_{\rm gy}^{-1}\delta_{\rm gc}^{3} \right\rangle,
\label{eq:Psi_phi}
\end{equation}
so that the gyrokinetic Poisson equation (\ref{eq:gyro_Poisson}) may be expressed in terms of the gyrokinetic variational principle
\cite{Brizard_VP, Brizard_gyroVP}
\begin{equation}
\frac{\delta}{\delta\phi} \left[ \int \frac{d^{3}r}{8\pi}\;|\nabla\phi|^{2} \;-\; \int d^{6}\ov{Z}\; \ov{F} \left( \frac{\ov{p}_{\|}^{2}}{2m} + \ov{\mu}\,B \;+\; q\,\Psi_{\rm gy}[\phi] \;-\; \ov{w} \right) \;\right] \;=\; 0.
\label{eq:Poisson_VP}
\end{equation}
The existence of a variational principle for the gyrokinetic Vlasov-Poisson and Vlasov-Maxwell equations allows us to compute exact conservation laws by Noether method \cite{Brizard_gyroVP,Brizard_Hahm}. Note that, according to the functional derivative (\ref{eq:Psi_phi}), the gyrocenter polarization effects (associated with the difference $\langle{\sf T}_{\rm gy}^{-1}\delta_{\rm gc}^{3}\rangle - \langle\delta_{\rm gc}^{3}\rangle$) require that quadratic nonlinearities in the electrostatic potential $\phi$ be retained in the effective gyrocenter potential (\ref{eq:Psi_gy}).

Next, by introducing the gyrocenter gyroradius vector
\begin{equation}
\vb{\rho}_{\rm gy} \;\equiv\; {\sf T}_{\rm gy}^{-1}\left( {\bf X} + \vb{\rho}_{\rm gc}\right) \;-\; \left( \ov{{\bf X}} + \vb{\rho}_{\rm gc} \right),
\label{eq:rho_gy_def}
\end{equation}
the push-forward expression (\ref{eq:push_delta}) may be written as ${\sf T}_{\rm gy}^{-1}\delta_{\rm gc}^{3} \equiv \delta^{3}(\ov{{\bf X}} + 
\vb{\rho}_{\rm gc} + \vb{\rho}_{\rm gy} - {\bf r})$. When expressed in terms of $(S_{1}, S_{2}, \cdots)$, the gyrocenter gyroradius vector (\ref{eq:rho_gy_def}) becomes
\begin{eqnarray}
\vb{\rho}_{\rm gy} & = & -\; \epsilon\,\left\{ S_{1},\; {\bf X} + \vb{\rho}_{\rm gc} \right\}_{\rm gc} \;-\; \epsilon^{2}\,\left\{ S_{2},\; {\bf X} + 
\vb{\rho}_{\rm gc} \right\}_{\rm gc} \nonumber \\
 &  &+\; \frac{\epsilon^{2}}{2}\;\left\{ S_{1},\frac{}{} \left\{ S_{1},\; {\bf X} + \vb{\rho}_{\rm gc} \right\}_{\rm gc} \right\}_{\rm gc} 
\;+\; \cdots \nonumber \\
 & \equiv & \ov{\vb{\rho}}_{\rm gy} \;+\; \wt{\vb{\rho}}_{\rm gy}, 
\label{eq:rho_gy}
\end{eqnarray}
where the gyroangle-dependent part of the gyrocenter gyroradius vector (\ref{eq:rho_gy}) is
\begin{equation}
\wt{\vb{\rho}}_{\rm gy} \;=\; -\,\epsilon\;\left\{ S_{1},\; {\bf X}\right\}_{\rm gc} \;+\; \cdots \;=\; \epsilon\;\frac{c\bhat}{B\Omega}\btimes\nabla
\wt{\Phi}_{\rm gc} \;+\; \cdots,
\label{eq:rhogyro_wt}
\end{equation}
while, up to second order, the gyroangle-independent part of the gyrocenter gyroradius vector (\ref{eq:rho_gy}) is
\begin{eqnarray}
\ov{\vb{\rho}}_{\rm gy} & = & -\;\epsilon\, \left\langle\left\{ S_{1},\; \vb{\rho}_{\rm gc}\right\}_{\rm gc}\right\rangle \;+\; \frac{\epsilon^{2}}{2}\,\left\langle \left\{ S_{1},\frac{}{} \left\{ S_{1},\; {\bf X}\right\}_{\rm gc} \right\}_{\rm gc} \right\rangle \label{eq:rhogyro_ov} \\
 & = & -\,\epsilon\;\frac{q}{B}\,\pd{}{\mu} \left\langle \vb{\rho}_{\rm gc} \left( \wt{\phi}_{\rm gc} \;-\; \frac{1}{\Omega}\,
\frac{d_{\rm gy}\wt{\Phi}_{\rm gc}}{dt} \right)\right\rangle \;-\; \epsilon^{2}\;\frac{q}{2\,\Omega^{2}}\left\langle \left\{ 
\wt{\Phi}_{\rm gc},\; \frac{c\bhat}{B}\btimes\nabla\wt{\Phi}_{\rm gc} \right\}_{\rm gc} \right\rangle, \nonumber
\end{eqnarray}
where corrections of order $\Omega^{-1}$ were kept. Expansion of the last term on the right side of the gyrokinetic Poisson equation 
(\ref{eq:gyro_Poisson}) in powers of $\vb{\rho}_{\rm gy}$ yields the expression
\begin{equation}
-\;\frac{\nabla^{2}\phi({\bf r})}{4\pi} \;=\; q\;\int d^{3}\ov{p}\,\ov{F} \;-\; \nabla\bdot\left( q\,\int d^{3}\ov{p}\;\ov{\vb{\rho}}_{\rm gy}\;
\ov{F} \;+\; \cdots \right),
\label{eq:Poisson_pol}
\end{equation}
where we have ignored FLR effects in the first term on the right side (i.e., $\langle\delta_{\rm gc}^{3}\rangle \rightarrow
\delta^{3}$), and the second term represents the polarization density. By keeping only terms of first order in $\epsilon$ and lowest FLR order, where
\[ \wt{\phi}_{\rm gc} \;\simeq\; \vb{\rho}_{\rm gc}\bdot\nabla\phi \;\;\;{\rm and}\;\;\; \wt{\Phi}_{\rm gc} \;\simeq\; -\,\vb{\rho}_{\rm gc}\bdot\bhat\btimes\nabla\phi, \]
the gyrorangle-independent gyrocenter gyroradius vector (\ref{eq:rhogyro_ov}) is expressed as
\begin{equation}
\ov{\vb{\rho}}_{\rm gy} \;\simeq\; \epsilon\;\frac{\bhat}{\Omega}\btimes \left( \frac{c\bhat}{B}\btimes\nabla\phi \;-\; \frac{c}{B\,\Omega}\;
\frac{d_{\rm gy}}{dt}\nabla\phi \right),
\label{eq:gyrorho_ZLR}
\end{equation}
which involves the E $\times$ B velocity and the polarization-drift velocity. 

Because of the mass dependence appearing in (\ref{eq:gyrorho_ZLR}), ion polarization effects in the gyrokinetic Poisson equation (\ref{eq:Poisson_pol}) dominate over electron polarization. Lastly, in the standard nonlinear gyrokinetic formalism \cite{Brizard_Hahm}, the polarization-drift contribution appears at second order in $\epsilon$ and is, therefore, omitted from the first-order gyrocenter gyroradius vector (\ref{eq:gyrorho_ZLR}). The standard gyrokinetic Poisson equation (\ref{eq:Poisson_pol}) thus yields the following relation between the electron (particle) density $n_{e}$ and the ion (gyrocenter) density $\ov{n}_{i}$:
\begin{equation}
e\,n_{e} \;=\; e\,\ov{n}_{i} \;+\; \nabla\bdot \left[\;\left( {\bf I} \;+\; \frac{\ov{n}_{i}c^{2}}{B^{2}/(4\pi\,m_{i})}\;{\bf I}_{\bot} \right)\bdot
\frac{\nabla\phi}{4\pi} \;\right],
\label{eq:Poisson_ei}
\end{equation}
where ${\bf I}_{\bot} = {\bf I} - \bhat\,\bhat$ is the perpendicular unit matrix. In (\ref{eq:Poisson_ei}), we note that the quasi-neutrality condition [i.e., the left side of (\ref{eq:Poisson_pol}) is zero] appears in the limit $B^{2}/(4\pi\,m_{i}\ov{n}_{i}) \ll c^{2}$. In this limit, 
(\ref{eq:Poisson_ei}) becomes the gyrokinetic quasi-neutrality condition
\begin{equation}
e\,n_{e} \;=\; e\,\ov{n}_{i} \;+\; \nabla_{\bot}\bdot\left( \frac{m_{i}c^{2}}{B^{2}}\;\ov{n}_{i}\;\nabla_{\bot}\phi\right),
\label{eq:gyro_quasi}
\end{equation}
which relates the electron (particle) density $n_{e}$, the ion (gyrocenter) density $\ov{n}_{i}$, and the electrostatic potential $\phi$ (through the ion polarization density). It is important to note that the ion (gyrocenter) density $\ov{n}_{i}$ must be defined as the moment of the full ion gyrocenter Vlasov distribution $\ov{F}_{i}$ [i.e., it is a solution of the nonlinear gyrokinetic Vlasov equation (\ref{eq:gyVE})] in order to conserve the global energy of the gyrokinetic Vlasov-Poisson equations \cite{Dimits, Dubin}.

\section{\label{sec:sum}Summary}

The guiding-center recursive Vlasov (gcrV) method yields results that are identical to the gyrocenter Lie-transform (gyLt) method. The Lie-transform method, however, offers several computational advantages. First, instead of computing each gyrocenter variable individually (gcrV method), the derivation of gyrocenter variables by gyLt method involves a single function $S_{k}$ at each order $\epsilon^{k}$ ($k = 1, 2, ...$) of the perturbation analysis. Moreover, we point out that, for most practical applications, the guiding-center and gyrocenter transformation can be kept separate since
\[ {\sf T}_{{\rm gc}}\,{\sf T}_{{\rm gy}}\,\ov{F} \;=\; {\ov F} \;+\; \left( \epsilon\;G_{1{\rm gy}}^{\alpha} \;+\; \epsilon_{B}\;
G_{1{\rm gc}}^{\alpha} \right) \partial_{\alpha}\ov{F} \;+\; \cdots. \]
Second, the explicit use of the guiding-center and gyrocenter pull-back and push-forward (Lie-transform) operators provides us with a simple interpretation of the recursive Vlasov method: the pull-back operator generates a fast-time-scale integration of the Vlasov dynamics while the push-forward operator represents the polarization dynamics in the gyrokinetic Poisson equation. Third, a self-consistent set of gyrokinetic Vlasov-Poisson
equations is obtained by the Lie-transform method since it is derived from a variational principle (which guarantees the existence of exact conservation laws). Lastly, the Lie-transform method can easily be generalized to the fully electromagnetic case \cite{Brizard_Hahm}.

%One considers the one-particle dynamics described in terms of the Poincar\'{e}-Cartan form. This ``abstract-mechanical'' approach allows to utilize the apparatus of the modern classical mechanics (in contrast with the iterative Vlasov approach where the tools available are limited to a basic perturbation calculus).

\acknowledgments

One of us (A.M.) wishes to acknowledge travel support from the EURATOM Association (Staff Movement under the Agreement on the Promotion
of Staff Mobility in the Field of Controlled Thermonuclear Fusion No.~131-83-7-FUSC/ERB 5005 CT 99 0080/ FU05 CT 2002 00010).

\appendix

\section{Guiding-center Transformation}

In a nonuniform magnetic field (where $\epsilon_{B} \equiv \rho/L_{B}$ denotes the dimensionless ratio of the characteristic gyroradius to the magnetic nonuniformity length scale), the guiding-center phase-space transformation is defined in terms of the first-order components
\begin{eqnarray}
G_{1}^{{\bf x}} & = & -\;\vb{\rho}_{0} \;\equiv\; -\;(2\,\mu B/m\Omega^{2})^{\frac{1}{2}}\;\wh{\rho}, \label{eq:gc_x} \\
G_{1}^{p_{\|}} & = & (mc/e)\,\mu \left( {\sf a}_{1}:\nabla\bhat \;+\; \bhat\bdot\nabla\btimes\bhat\right) \;-\; p_{\|}\;\vb{\rho}_{0}\bdot\left(\bhat\bdot\nabla\bhat\right), \label{eq:gc_p} \\
G_{1}^{\mu} & = & \vb{\rho}_{0}\bdot \left( \mu\;\nabla\ln B \;+\; \frac{mv_{\|}^{2}}{B}\;\bhat\bdot\nabla\bhat \right) \;-\; \mu\;\frac{v_{\|}}{\Omega} \left( {\sf a}_{1}:\nabla\bhat \;+\; \bhat\bdot\nabla\btimes\bhat \right), \label{eq:gc_mu} \\
G_{1}^{\theta} & = & -\;\vb{\rho}_{0}\bdot{\bf R} \;+\; \pd{\vb{\rho}_{0}}{\theta}\bdot\nabla\ln B \;+\; 
\frac{v_{\|}}{\Omega}\;{\sf a}_{2}:\nabla\bhat \;+\; \frac{mv_{\|}^{2}}{2\,\mu B}\;\left( \bhat\bdot\nabla\bhat\bdot\pd{\vb{\rho}_{0}}{\theta} \right) \label{eq:gc_theta}
\end{eqnarray}
where we used the definitions for the gyroangle-dependent rotating unit vectors $\wh{\rho} \equiv \cos\theta\,\wh{{\sf e}}_{1} - \sin\theta\,
\wh{{\sf e}}_{2}$ and $\wh{\bot} \equiv {\bf v}_{\bot}/|{\bf v}_{\bot}| = \partial\wh{\rho}/\partial\theta$, expressed in terms of the fixed unit vectors $(\wh{{\sf e}}_{1}, 
\wh{{\sf e}}_{2}, \bhat \equiv \wh{{\sf e}}_{1}\btimes\wh{{\sf e}}_{2})$, so that the gyrogauge vector ${\bf R} = \nabla\wh{\bot}\bdot\wh{\rho} \equiv 
\nabla\wh{{\sf e}}_{1}\bdot\wh{{\sf e}}_{2}$ is gyroangle independent and the dyadic (traceless) tensors 
\[ {\sf a}_{1} \;=\; -\,\frac{1}{2}\,(\wh{\rho}\,\wh{\bot} + \wh{\bot}\,\wh{\rho}) \;=\; \frac{1}{2} \left[ \left(\wh{{\sf e}}_{1}\wh{{\sf e}}_{1}
- \wh{{\sf e}}_{2}\wh{{\sf e}}_{2}\right)\; \sin\,2\theta \;+\frac{}{} \left(\wh{{\sf e}}_{1}\wh{{\sf e}}_{2} + \wh{{\sf e}}_{2}\wh{{\sf e}}_{1}\right)\; 
\cos\,2\theta \right] \]
and ${\sf a}_{2} \equiv \int{\sf a}_{1}\,d\theta$ are gyroangle dependent. In (\ref{eq:gc_x})-(\ref{eq:gc_theta}), $\vb{\rho}_{\rm gc} \equiv 
{\sf T}_{\rm gc}^{-1}{\bf x} - {\bf X} = \vb{\rho}_{0} + \epsilon_{B}\,\vb{\rho}_{1} + \cdots$ denotes the guiding-center gyroradius, where 
${\sf T}_{\rm gc}^{-1}$ denotes the guiding-center push-forward operator (see below) and $\vb{\rho}_{1}$ is the first-order correction to the lowest-order gyroradius vector $\vb{\rho}_{0}$.

In a nonuniform magnetic field, the guiding-center pull-back operator is expressed as
\begin{equation}
{\sf T}_{{\rm gc}} \;\equiv\; \exp \left[ -\,\vb{\rho}_{0}\bdot\nabla + \epsilon_{B}\,\left( G_{1}^{p_{\|}}\;\pd{}{p_{\|}} + G_{1}^{\mu}\;
\pd{}{\mu} + G_{1}^{\theta}\;\pd{}{\theta} + G_{2}^{{\bf x}}\bdot\nabla \right) + \cdots \right],
\label{eq:pullback_gc_exact}
\end{equation}
and the guiding-center push-forward ${\sf T}_{\rm gc}^{-1}$ can be used to construct the guiding-center gyroradius vector
\begin{equation}
\vb{\rho}_{\rm gc} \;=\; \vb{\rho}_{0} \;+\; \epsilon_{B} \left[ \frac{1}{2}\,\vb{\rho}_{0}\bdot\nabla\vb{\rho}_{0} \;-\; \frac{1}{2} \left(
G_{1}^{\mu}\;\pd{\vb{\rho}_{0}}{\mu} + G_{1}^{\theta}\;\pd{\vb{\rho}_{0}}{\theta} \right) \;-\; G_{2}^{{\bf x}} \right] \;+\; \cdots,
\label{eq:rhogc_def}
\end{equation} 
where $\vb{\rho}_{0}\bdot\nabla\vb{\rho}_{0} = -\,\frac{1}{2}\,(\vb{\rho}_{0}\bdot\nabla\ln B)\;\vb{\rho}_{0} - (\vb{\rho}_{0}\bdot\nabla\bhat\bdot
\vb{\rho}_{0})\,\bhat - (\vb{\rho}_{0}\bdot{\bf R})\,\partial\vb{\rho}_{0}/\partial\theta$. 

The Jacobian for the guiding-center transformation is expressed (up to first order in $\epsilon_{B}$) as
\begin{eqnarray}
{\cal J}_{{\rm gc}} & \equiv & {\cal J} \;-\; \epsilon_{B}\,\pd{}{Z^{\alpha}} \left( {\cal J}\;G_{1}^{\alpha} \right) \nonumber \\
 & = & B \;+\; \epsilon_{B}\,\nabla\bdot(B\,\vb{\rho}_{0}) \;-\; \epsilon_{B}\,B \left( \pd{G_{1}^{p_{\|}}}{p_{\|}} \;+\; \pd{G_{1}^{\mu}}{\mu} \;+\; \pd{G_{1}^{\theta}}{\theta} \right) \nonumber \\
 & = & B \left( 1 \;+\; \epsilon_{B}\,\frac{v_{\|}}{\Omega}\;\bhat\bdot\nabla\btimes\bhat \right) \;\equiv\; B_{\|}^{*}.
\label{eq:Jac_gc}
\end{eqnarray}
The guiding-center Poisson bracket, on the other hand, is expressed in terms of two arbitrary functions $F$ and $G$ as
\begin{eqnarray}
\{ F,\; G\}_{\rm gc} & = & \left( \pd{F}{w}\,\pd{G}{t} - \pd{F}{t}\,\pd{G}{w} \right) \;+\; \frac{\Omega}{B} \left( \pd{F}{\theta}\,\pd{G}{\mu} - \pd{F}{\mu}\,\pd{G}{\theta} \right) \nonumber \\
 &  &+\; \frac{{\bf B}^{*}}{B_{\|}^{*}}\bdot\left( \nabla F\,\pd{G}{p_{\|}} - \pd{F}{p_{\|}}\,\nabla G \right) \;-\; \frac{c\bhat}{q\,B_{\|}^{*}}\bdot\left(\;\nabla F\btimes\nabla G\frac{}{}\right),
\label{eq:gcPB_general}
\end{eqnarray}
where ${\bf B}^{*} \equiv {\bf B} + \epsilon_{B}\,(p_{\|}c/q)\,\nabla\btimes\bhat$ and $B_{\|}^{*} \equiv \bhat\bdot{\bf B}^{*}$.

Lastly, the guiding-center pull-back operator (\ref{eq:pullback_gc_exact}) and the guiding-center Jacobian (\ref{eq:Jac_gc}) can be used to obtain the push-forward representation of the particle Vlasov-moment integral
\begin{equation}
\|\chi\| \;=\; \int d^{6}z\,{\cal J}\;\chi\,\delta^{3}({\bf x} - {\bf r})\;{\sf T}_{{\rm gc}}F \;=\; \int d^{6}Z\,{\cal J}_{{\rm gc}}\;
{\sf T}_{{\rm gc}}^{-1}\chi\; \delta^{3}({\bf X} + \vb{\rho}_{\rm gc} - {\bf r})\;F,
\label{eq:chi}
\end{equation} 
where $\chi$ is an arbitrary function in particle phase space and we used the identity
\[ {\cal J}\,S \;-\; \epsilon_{B}\,\pd{}{Z^{\alpha}} \left( {\cal J}\,G_{1}^{\alpha}\;S \right) \;+\; \cdots \;=\; {\cal J}_{{\rm gc}}\,\left( S \;-\; 
\epsilon_{B}\,G_{1}^{\alpha}
\;\pd{S}{Z^{\alpha}} \;+\; \cdots \right) \;\equiv\; {\cal J}_{{\rm gc}}\,{\sf T}_{{\rm gc}}^{-1}S. \]
If we now expand the delta function $\delta^{3}({\bf X} + \vb{\rho}_{\rm gc} - {\bf r})$ in powers of $\vb{\rho}_{\rm gc}$ and integrate by parts, we obtain the guiding-center push-forward representation
\begin{equation}
\|\chi\| \;\equiv\; \left\|{\sf T}_{{\rm gc}}^{-1}\chi\right\|_{{\rm gc}} \;-\; \nabla\bdot\left\|\vb{\rho}_{\rm gc}\;{\sf T}_{{\rm gc}}^{-1}\chi
\right\|_{{\rm gc}} \;+\; \cdots,
\label{eq:chi_gc}
\end{equation}
which enables us to write particle fluid moments in terms of guiding-center fluid moments.

\end{document}